\documentclass[12pt,preprint]{aastex}

\shorttitle{A deep, high resolution survey of the low frequency radio sky}
\shortauthors{Lenc Garrett Wucknitz Anderson Tingay}

\newcommand\rah{\mbox{$^{\mathrm h}$}}%
\newcommand\ram{\mbox{$^{\mathrm m}$}}%

\begin{document}

\title{A deep, high resolution survey of the low frequency radio sky}

\author{E. Lenc}

\affil{Centre for Astrophysics and Supercomputing, Swinburne University of Technology,
Mail number H39, P.O. Box 218, Hawthorn, Victoria 3122}
\email{elenc@astro.swin.edu.au}

\author{M.A. Garrett\altaffilmark{1}}
\affil{Netherlands Foundation for Research in Astronomy (ASTRON), Postbus 2, 7990 AA Dwingeloo, The Netherlands}
\email{garrett@astron.nl}

\author{O. Wucknitz\altaffilmark{2}, J.M. Anderson}
\affil{Joint Institute for VLBI in Europe, Postbus 2, 7990 AA Dwingeloo, The Netherlands}
\email{wucknitz@jive.nl, anderson@jive.nl}

\and

\author{S.J. Tingay}
\affil{Department of Imaging and Applied Physics, Curtin University of Technology, Bentley 6845, Western Australia, Australia}
\email{}

\altaffiltext{1}{Adjunct Professor, Centre for Astrophysics and Supercomputing, Swinburne University of Technology, Mail number H39, P.O. Box 218, Hawthorn, Victoria 3122}
\altaffiltext{2}{Argelander-Institut f\"ur Astronomie, Universit\"at Bonn, Auf dem H\"ugel 71, 53121, Bonn, Germany; wucknitz@astro.uni-bonn.de}
\begin{abstract}

We report on the first wide-field, very long baseline interferometry (VLBI) survey at 90 cm.  The survey area consists of two overlapping 28 deg$^{2}$ fields centred on the quasar J0226$+$3421 and the gravitational lens B0218$+$357. A total of 618 sources were targeted in these fields, based on identifications from Westerbork Northern Sky Survey (WENSS) data. Of these sources, 272 had flux densities that, if unresolved, would fall above the sensitivity limit of the VLBI observations. A total of 27 sources were detected as far as $2\arcdeg$ from the phase centre. The results of the survey suggest that at least $10\%$ of moderately faint (S$\sim100$ mJy) sources found at 90 cm contain compact components smaller than $\sim0.1$ to $0.3$ arcsec and stronger than $10\%$ of their total flux densities. A $\sim90$ mJy source was detected in the VLBI data that was not seen in the WENSS and NRAO VLA Sky Survey (NVSS) data and may be a transient or highly variable source that has been serendipitously detected. This survey is the first systematic (and non-biased), deep, high-resolution survey of the low-frequency radio sky. It is also the widest field of view VLBI survey with a single pointing to date, exceeding the total survey area of previous higher frequency surveys by two orders of magnitude. These initial results suggest that new low frequency telescopes, such as LOFAR, should detect many compact radio sources and that plans to extend these arrays to baselines of several thousand kilometres are warranted.

\end{abstract}


\section{Introduction}
\label{sec:introduction}

The general properties of the 90 cm sky are not very well known and even less is known at VLBI resolution. Previous snapshot surveys at these wavelengths have only targeted the brightest sources and were plagued by poor sensitivity, radio interference and limited coherence times. Furthermore, the field of view that could be imaged was typically limited by the poor spectral and temporal resolution of early generation hardware correlators and the available data storage and computing performance at the time. As a result, although several hundred 90 cm VLBI observations have been made over the past two decades, images of only a few tens of sources have been published e.g. \citet{alt95}; \citet{laz98}; \citet{chu99}; \citet{cai02}. With such a small sample it is difficult to quantify the total population and nature of these sources. In particular, the sub-arcsecond and sub-Jansky population of 90 cm sources is largely unexplored. 

Recent improvements to the EVN hardware correlator at JIVE \citep{van04}, have enabled significantly finer temporal and spectral resolution. Combined with vast improvements in storage and computing facilities, it is now possible to image fields as wide as, or even wider than, the FWHM of the primary beam of the observing instrument. To complement the hardware improvements, new approaches to calibration and imaging have been developed to better utilise the available data and processing platforms. For example, \cite{gar05} performed a deep VLBI survey at 20 cm of a $36\arcmin$ wide field by using a central bright source as an in-beam calibrator. The approach was ideal for survey work as it permitted the imaging of many potential target sources simultaneously by taking advantage of the full sensitivity of the observation across the entire field of view. We have applied a similar technique at 90 cm by piggybacking on an existing VLBI observation of the gravitational lens B0218$+$357 and the nearby quasar J0226$+$3421, with the aim of surveying a 28 deg$^{2}$ field around each of the sources. The results provide an important indication of what may be seen by future low-frequency instruments such as the Low Frequency Array (LOFAR), European LOFAR (E-LOFAR) and the Square Kilometre Array (SKA).

In this paper, we present the results of a 90 cm wide-field VLBI survey that covers two partially overlapping regions of 28 deg$^{2}$ each, surveying 618 radio source targets at angular resolutions ranging between 30 and 300 mas. For sources located at a redshift of $z=1$, the linear resolution corresponding to 30 mas is 230 pc. A \emph{WMAP} cosmology \citep{spe03} with a flat Universe, $H_{0}=72$ km s$^{-1}$ Mpc$^{-1}$ and $\Omega_{m}=0.29$ is assumed throughout this paper.

\section{Observations and Correlation}

A VLBI observation of the gravitational lens B0218$+$357 was made on 11 November 2005 using all ten NRAO Very Long Baseline Array (VLBA) antennas, the Westerbork Synthesis Radio Telescope (WSRT) as a phased array and the Jodrell Bank, 76$-$m Lovell Telescope (JB). The primary aim of this observation was to investigate, in detail, propagation effects in the lensing galaxy and the substructure in the lens. The secondary aim, as investigated in this paper, was to serve as a wide-field test observation to study the faint source population at 90 cm over a good fraction of the primary beam.

The observation spanned 14 hours with approximately 6 hours of data recorded at JB and WSRT and 13 hours at the VLBA stations. Ten minutes scans of the target source B0218$+$357 ($\alpha=02\rah21\ram05\fs4733$ and $\delta=35\arcdeg56\arcmin13\farcs791$) were interleaved with three minute scans of the nearby quasar J0226$+$3421 ($\alpha=02\rah26\ram10\fs3332$ and $\delta=34\arcdeg21\arcmin30\farcs286$). Five minute scans of the fringe finder 3C84 were made approximately every four hours. Dual circular and cross polarisation data were recorded across four 4 MHz IFs centred on 322.49, 326.49, 330.49 and 610.99 MHz respectively, resulting in a total data recording rate of 128 Mbits s$^{-1}$. The 610.99 MHz data were only recorded at the VLBA antennas. The data were correlated at the European VLBI Network (EVN) correlator at the Joint Institute for VLBI in Europe (JIVE, Dwingeloo, the Netherlands) in multiple passes to create a single-IF, single polarisation, wide-field data set and a multi-IF, dual circular polarisation, narrow-field data set. The narrow-field data set concentrated on the B0218$+$357 with greater sensitivity and the results of this observation will be presented elsewhere (Wucknitz et al., in preparation). The wide-field data consisted of a single polarisation (LL), single IF with a 4 MHz band centred on 322.49 MHz. A third correlator pass centred on another source was used to create a second wide-field data set with a single IF and RR polarisation but was not used in our data reduction process.

To reduce the effects of bandwidth smearing and time averaging smearing, and thus image the largest possible field, the EVN correlator generated data with 512 spectral points per baseline and an integration time of 0.25 s. The spectral and temporal resolution exercised the current physical limits of the JIVE hardware correlator and resulted in a final data set size of 77.5 Gbytes. The wide-field data set has a one sigma theoretical thermal noise of $\sim$1.2 and $\sim$0.7 mJy/beam for the quasar and gravitational lens, respectively.

\section{VLBI Calibration and Imaging}
\label{sec:vlbical}
The data from the narrow-field data set were used to perform the initial editing and calibration of the phase reference to take advantage of the increased sensitivity available with the additional bands and polarisations. Nominal corrections to counter the effects of the total electron content (TEC) of the ionosphere were applied with the AIPS\footnote{The Astronomical Image Processing System (AIPS) was developed and is maintained by the National Radio Astronomy Observatory, which is operated by Associated Universities, Inc., under co-operative agreement with the National Science Foundation} task TECOR. An amplitude calibration table was derived from measures of the system temperature of each antenna throughout the observation using the AIPS task APCAL, and applied to the data set.

Delays across the IF bands, which were assumed to be constant throughout the observation, were calibrated by fringe fitting on 3C84. A multi-band fringe fit was then performed on the quasar. The flagging and calibration tables of the narrow-field data set were transferred to the wide-field data set using a ParselTongue\footnote{A Python scripting tool for AIPS. ParselTongue was developed in the context of the ALBUS project, which has benefited from research funding from the European Community's sixth Framework Programme under RadioNet R113CT 2003 5058187. ParselTongue is available for download at \url{http://www.radionet-eu.org/rnwiki/ParselTongue}} script. Further editing was applied to the wide-field data set to remove the frequency band edges and frequency channels adversely affected by RFI. The bandpass for the observation was calibrated against observations of 3C84.

As both the phase reference and the gravitational lens were to be used as in-beam calibrators for their respective fields, accurate calibration of the amplitudes and phases of both fields was essential. The calibration was complicated by the complex structure of both sources. To account for this structure a new DIFMAP \citep{she94} task, \emph{cordump}\footnote{The \emph{cordump} patch is available for DIFMAP at \url{http://astronomy.swin.edu.au/$\sim$elenc/DifmapPatches/}} \citep{len06}, was developed to enable the transfer of all phase and amplitude corrections made in DIFMAP during the imaging process to an AIPS compatible SN table. The \emph{cordump} task greatly simplified the calibration of both of the fields. First, the phase reference data were averaged in frequency and exported to DIFMAP where several iterations of modelling and self-calibration of both phases and amplitudes were performed. \emph{cordump} was then used to transfer the resulting phase and amplitude corrections back to the unaveraged AIPS data set. After application of these corrections, the DIFMAP model of the phase reference source was subtracted from the AIPS (u-v) data set. The quasar self-calibration solutions were then applied to the lens field as an initial calibration for that field. The calibration for the lens field was further refined using the same approach as for the quasar and upon completion the DIFMAP model of the lens was subtracted from the calibrated field. The images of the phase reference and the lens had measured RMS noise of 1.8 and 1.0 mJy beam$^{-1}$, respectively. The higher than theoretical noise is attributed to substantial levels of RFI observed on some baselines and the shorter time available with the WSRT and JB observations.

In the first phase of the imaging process, the AIPS task IMAGR was used to make naturally weighted dirty images and beams of regions selected from WENSS data \citep{ren97} of the two fields being surveyed, the source selection criteria are described in detail in \S~\ref{sec:selection}. Targets falling within a certain annulus around each field were imaged simultaneously using the multi-field option within IMAGR, the DO3D option was used to reduce non-coplanar array distortion. The data from both fields were kept in an unaveraged form to prevent smearing effects during imaging. For each target, the dirty image subtended a square of approximately $51\arcsec$ on each side, an area that covers approximately half that of the WENSS beam at the observation declination ($54\arcsec\times92\arcsec$). Since the dirty image of each target source contains $\sim2\times10^5$ synthesized-beam areas, a conservative $6\sigma$ detection threshold was imposed to avoid spurious detections. Furthermore, only the inner 75\% of each dirty image was searched for candidate detections to avoid erroneous detections as a result of map edge effects. For each positive detection the coordinate of the VLBI peak flux density was recorded. This first imaging step was used to determine whether the target source had been detected with the VLBI observation.  Based on our detection criteria, we estimate a false detection rate of approximately one in every 3300 images. While we expect the majority of unresolved WENSS sources to have peaks that fall within the imaged areas, approximately 9.5\% of the WENSS sources exhibit resolved structure. For these sources, we would not detect bright compact components that may exist outside of the central region that was imaged.

During the calibration process, it was noted that the lens had significantly weaker signal on the longer baselines compared to that of the quasar. To test the effectiveness of the refined lens field self-calibration solutions, we re-imaged one of the B0218$+$357 field sources, B0215.1$+$3710, with only the phase reference self-calibration solutions applied. B0215.1$+$3710 is located on the side of the lens field that is furthest from the phase reference ($\sim3.45\arcdeg$) and so is most sensitive to changes from the nominal conditions that were corrected for. The target-calibrated dirty image for this source has an rms noise of 4.7 mJy beam$^{-1}$ and a peak of 42 mJy beam$^{-1}$. With only the phase reference self-calibration solutions applied, the rms noise is 6.6 mJy beam$^{-1}$ and the peak 21 mJy beam$^{-1}$. It is clear that without the refined calibration, B0215.1$+$3710 would not have been detected above the $6\sigma$ threshold.

The second phase of the imaging process involved creating a (u,v) shifted data set for each of the positive detections, using the AIPS task UVFIX, such that the new image centre coincided with the coordinate of the image peak recorded in the first phase. The shifted data sets were averaged in frequency, effectively reducing the field of view of each of the targeted sources to approximately $0.5\arcmin$, and then exported to DIFMAP. In DIFMAP, the visibilities were averaged over 10 second intervals to reduce the size of the data set and to speed up the imaging process. Each target was imaged in DIFMAP, with natural weighting applied, using several iterations of model fitting. Phase self-calibration was performed between iterations to adjust for the varying effects of the ionosphere across the field.

During the imaging process it was noted that the self-calibration phase corrections varied significantly between fields and even among sources within each field. It is believed that these were due to ionospheric variations that occurred across the survey field. Observations at 90 cm will invariably suffer degradation as a result of ionospheric variations and the nominal TEC corrections made in the initial calibration stages assumed that these corrections would be valid across the entire field. This is not a valid assumption when imaging extremely large fields. To provide position dependent corrections within a field, ParselTongue scripts were developed to implement two alternate methods that could be tested against the data.

The first method calculated ionospheric corrections based on TEC measures to each source of interest in the field of view and applied differential corrections, based on the TEC correction already made at the phase centre, prior to imaging. This allowed a differential correction to be applied after self calibrating on the bright central sources in each field of view of these observations. The corrections did not result in any significant improvement in the resulting images. We suspect that the currently available TEC solutions may be too coarse, both spatially and temporally, to account for the ionospheric variations across the field.

The second method used a parameterized ionospheric model to determine the corrections to each source of interest in the field of view. As with the TEC corrections, these were applied after the central source in each field had been self calibrated. Preliminary tests of these corrections indicated that they performed better than the differential TEC solutions with improvements of the order of a few percent in flux density observed in approximately 70\% of sources tested.

The testing of these libraries is not yet complete and only the 11 wide-field sources of \citet{len06} were used in our initial tests. Further tests will be required to more robustly analyse the performance of the two approaches to ionospheric calibration.

Following our first attempt to survey the inner $0\arcdeg-1\arcdeg$ region of each field \citep{len06} we discovered that the positional accuracy of the detected sources degraded significantly with radial distance from the phase centre when compared to the positions derived from observations with other instruments. While most sources observed with other instruments only had a positional accuracy of $\sim1\arcsec$ it was still clear that our fields were being scaled by a factor of $0.99871\pm7\times10^{-5}$, a factor that corresponds to an offset of $53\pm3$ frequency channels in our data set. Interestingly, this appeared to corresponded to the 50 lower-band channels that were flagged during editing in the 31DEC05 version of AIPS that was used for the processing of the data. When the processing was repeated in the 31DEC06 version of AIPS, the positional discrepancies disappeared so we suspected that there may have been a software issue with the earlier version of AIPS. However, as wider fields were imaged we discovered four sources that were common to both of the fields. Each of these sources should have been well aligned between the two fields, however, discrepancies of $0.41\arcsec-0.65\arcsec$ were being observed in approximately the same position angle. This was also indicative of a radial scaling but to a lesser degree, $0.999927\pm1.4\times10^{-5}$, corresponding to an offset of $3\pm0.6$ channels. The source of this error has not yet been identified, however all of the sources positions and images in this paper have been corrected to account for this effect.

\section{Survey annuli, survey depths, and source selection}
\label{sec:selection}
We split the survey of each field into six annuli based on the radial distance from the correlation phase centre of that field. These are referred to as the $0\arcdeg-0.25\arcdeg$, $0.25\arcdeg-0.5\arcdeg$, $0.5\arcdeg-1\arcdeg$, $1\arcdeg-1.5\arcdeg$, $1.5\arcdeg-2\arcdeg$, and $2\arcdeg-3\arcdeg$ annuli in field 1 (centred on J0226$+$3421) and field 2 (centred on B0218$+$357). Our survey attempts to detect sources at large radial distances from the antenna pointing position and the correlation phase centre. To reduce the effect of bandwidth and time-averaging smearing, increasingly restrictive (u,v) ranges are employed in the outer annuli. Furthermore, the fall-off of the response of the primary beam is an effect that significantly limits the sensitivity within each annulus. In particular, WSRT and JB have significantly narrower primary beams ($\sim1\arcmin$ and $\sim0.5\arcdeg$ HWHM respectively), owing to their larger effective aperture, compared to the VLBA ($\sim1.3\arcdeg$ HWHM). As such, the WSRT data were only used to image the source directly at the phase centre, whilst the JB observatory data was only used in the $0\arcdeg-0.25\arcdeg$ annulus (restrictions in (u,v) range effectively excludes the JB data from the $0.25\arcdeg-0.5\arcdeg$ annulus even though it has a significant response within this annulus).

Since only the VLBA antennas were used outside the $0\arcdeg-0.25\arcdeg$ annulus, the reduced response in the other annuli is composed of only three independent components: the VLBA primary beam response ($R_{VLBA}$) and the reduced response due to bandwidth and time-averaging smearing ($R_{bw}$, $R_{t}$). The combined reduced response, $R$, is given by $R=R_{bw}R_{t}R_{VLBA}$. We have estimated $R_{bw}$ and $R_{t}$ following \citet{bri99} and have adopted a fitted function used to model the VLA antennas, as documented in the AIPS task PBCOR, to model the primary beam response of the VLBA under the assumption that the 25 m antennas in these arrays have a similar response \citep{gar05}. In Table \ref{tab:tabfields}, we calculate the total response and estimated $1\sigma$ rms noise at the outer edge of each annulus of the two surveyed fields. The (u,v) range has been restricted to limit the effects of bandwidth and time-averaging smearing to at most a few percent.

The WENSS catalogue was used as a guide for potential targets in the survey. Table \ref{tab:tabfields} lists the total number of WENSS sources that exist within each annulus of each field. For a WENSS source to be detected by our survey it must have a peak flux density, $S_{P}($WENSS$)$, that satisfies the constraint $S_{P}($WENSS$)>6\sigma R^{-1}$. Estimates of this limit at the edge of each annulus, $S_{P}$, and the number of WENSS sources that meet this constraint, $\langle N_{VLBI}\rangle$, are listed in Table \ref{tab:tabfields}. Even though it was estimated that many of the WENSS sources would fall below our detection limits, for completeness, we targeted all WENSS sources in the $0\arcdeg-2\arcdeg$ annuli of each field given the possibility that some sources might exhibit strong variability. Between $2\arcdeg-3\arcdeg$ only candidate sources that were within our sensitivity limits were targeted.

\section{Results}

A total of 618 WENSS sources were targeted by the 90 cm wide-field VLBI survey at radial distances of up to $2.89\arcdeg$ from the phase centre of the survey fields. The complete survey of all WENSS sources within the inner $0\arcdeg-2\arcdeg$ annulus of each field did not detect any source that had peak flux below our sensitivity limit, or that was partially resolved in WENSS. The combined total area imaged around each of the targeted sources in these fields represents $\sim0.5\%$ of the area surveyed.

Of all of the WENSS sources targeted, a total of 272 sources, 95 in the J0226$+$3421 field (field 1) and 177 in the more sensitive B0218$+$357 field (field 2), would have peak flux densities above our VLBI detection limits, if unresolved by our VLBI observations. The WENSS characteristics of these sources (position, peak flux density per solid beam angle, integrated flux and where available, the WENSS/NVSS spectral index $\alpha$ where $S_{\nu}\propto\nu^{\alpha}$) are listed in Tables \ref{tab:tabf1} and \ref{tab:tabf2} for fields 1 and 2, respectively. Where the target source is also detected by the VLBI survey, the corrected source position (see \S~\ref{sec:vlbical}), VLBI peak flux density per solid beam angle and integrated flux are listed in the following row. The VLBI peak and integrated flux density have been corrected for the primary beam response but not for bandwidth and time-averaging smearing losses. Based on these losses and uncertainties in the amplitude calibration, we estimate the absolute flux density scales of the VLBI observations to be better than $10\%$ in the inner $0.5\arcdeg$ of each field and better than $20\%$ elsewhere.

A total of 27 sources were detected and imaged by the survey, eight of these sources detected in the J0226$+$3421 field (field 1) and the remaining 19 detected in the more sensitive B0218$+$357 field (field 2). Nine of the sources were detected outside of the half-power point of the VLBA primary beam (HWHM $\sim1.3\arcdeg$). Four sources, B0223.1$+$3408, B0221.9$+$3417, B0219.2$+$3339 and B0223.5$+$3542, were detected in both fields in the region of overlap. B0219.2+3339, in particular, was detected at $2.06\arcdeg$ from the phase centre of the second field and is well past the quarter power point of the VLBA primary beam, it is also the only source detected in the outer annulus of the survey. Table \ref{tab:tabastrometry} lists all of the detected sources, the detection field, the distance from the phase centre of that field, the restoring beam size and the one sigma residual RMS noise. A positional comparison of our corrected source positions is also made with respect to the best known radio positions, where $d_{E}$ and $\theta_{E}$ are the observed offset and position angle from this position, and $d_{\sigma}$ is the offset in terms of the combined one sigma position error of the two compared positions.

The VLBI positions in the $0\arcdeg-0.25\arcdeg$ and $0.25\arcdeg-0.5\arcdeg$ annuli are limited by errors introduced by the ionosphere.  We estimate a one sigma error of $\sim3-12$ mas in each coordinate. The positional accuracy of the outer, heavily tapered, annuli are further limited by RMS noise errors. We estimate the one sigma error in each coordinate of these outer fields to be better than 15 mas, 20 mas and 30 mas for the $0.5\arcdeg-1\arcdeg$, $1\arcdeg-1.5\arcdeg$ and $1.5\arcdeg-2\arcdeg$ annuli, respectively. The positional accuracy of B0219.2+3339 is also expected to be $\sim30$ mas as it is located quite close to the $2\arcdeg$ boundary. After a correction was applied for an apparent scaling effect in the image (see \S~\ref{sec:vlbical}), residual errors of 40 mas, 100 mas, 200 mas and 180 mas were measured for the cross-field detections of sources B0223.1$+$3408, B0221.9$+$3417, B0219.2$+$3339 and B0223.5$+$3542 respectively. These extremely wide-field sources exhibited significant ionospheric phase fluctuations that distorted their initial dirty images. While the fluctuations were partially corrected for with phase self-calibration, it is believed that this may have adversely affected the position accuracy.

Contour maps for each of the sources detected in field 1 are shown in Figure \ref{fig:figf1} and those detected in field 2 are shown in Figure \ref{fig:figf2}. While not formally part of the wide-field survey, the fringe finder 3C84 has also been imaged and is shown in Figure \ref{fig:fig3c84}.

The total survey required nearly six weeks of processing on a single 2 GHz computer, required $\sim200$ gigabytes of workspace, and generated a total of $\sim20$ gigabytes of image data. We estimate that six years of processing would be required to completely image the FWHM beam of the VLBA at 320 MHz using similar techniques. Fortunately, the problem can be easily broken down to run efficiently in the parallel environment of a supercomputing cluster. With a basic 100-node cluster, the entire FWHM beam of the VLBA could be imaged within three weeks using a simple brute force method that would image targets on individual processors. For a single field, the cluster would generate mosaic of the primary beam comprising of $\sim1$ terabyte of image data. More elaborate algorithms may be employed to improve the efficiency of this processing further, for example, by using a recursive approach that creates successively smaller sub-fields by performing a combination of (u-v) shifting and data averaging.

\subsection{Comments on individual sources}

\subsubsection{3C84}
For 3C84, we measure a VLBI peak flux density of 2.33 Jy beam$^{-1}$ and an integrated flux of 6.07 Jy, whereas the WENSS peak flux density and integrated flux is 19.396 Jy beam$^{-1}$ and 42.8 Jy respectively. This is the only source in our sample of imaged sources with extended structure in WENSS, where it has an estimated size of $115\arcsec\times84\arcsec$ at a position angle of $115\arcdeg$. In our VLBI image we have recovered $\sim14\%$ of the WENSS flux. Similar VLBI observations at 327 MHz, with a larger synthesised beam, measure a slightly greater integrated flux of 7.47 Jy \citep{ana89} suggesting that the missing flux is most likely related to the larger-scale structure that is resolved out by VLBI observations. We measure a Largest Angular Size (LAS) of 150 mas which corresponds to a Largest Linear Size (LLS) of $\sim50$ pc at its measured redshift of $z=0.017559\pm0.000037$ \citep{str92}. A 15 GHz VLBA contour map \citep{lis05} is shown overlaid with our 90 cm image of 3C84 in Figure \ref{fig:fig3c84}. The smaller scale structures within this image appear to align with the jet-like feature that appears in our 90 cm image and extends 100 mas to the south of the core.

\subsubsection{B0223.1+3408 (J0226$+$3421, 4C$+$34.07)}
The quasar, J0226$+$3421, was imaged with the full (u,v) range and recovers approximately 80\% of the WENSS flux. The contour map of this source is shown in Figure \ref{fig:figf1}(a) and is overlaid with a naturally weighted 2 cm A$-$configuration VLA with Pie Town contour map (Wucknitz et al., in preparation). The source is dominated by a bright core ($\sim0.85$ Jy) and an extended lobe to the west ($\sim1.6$ Jy). A weaker lobe appears to the north ($\sim0.28$ Jy) and a partially resolved hot spot ($\sim0.14$ Jy) approximately mid-way between the core and the western lobe. The source has a LAS of $1.15\arcsec$ which corresponds to a LLS of $\sim9$ kpc at its measured redshift of $z=2.91\pm0.002$ \citep{wil98}. All of the large-scale structures observed at 90 cm with VLBI are also clearly detected at 2 cm with the VLA and Pie Town. MERLIN$+$VLBI images at 18 cm \citep{dal95} detect the core and western lobe with large-scale structure and positions that are consistent with our image, however, their observations do not detect the northern lobe or hot spot.   The source was also detected in the second field and is shown in Figure \ref{fig:figf2}(q) with an image of the field 1 source restored using the same beam. An offset of 40 mas is observed between the field 1 and field 2 source after correcting for the larger-scale offset described in \S~\ref{sec:vlbical}.

\subsubsection{B0218.0+3542 (B0218$+$357)}
B0218.0$+$3542 is a gravitational lens that has been mapped at higher frequencies \citep[e.g.][]{big01,wuc04}, with VLBI \citep[e.g.][]{big03}, and at various wavelengths by \citet{mit06}. The source is the smallest known Einstein radio ring \citep{pat93}. As this source was the main target of the original observation, it is placed in the most sensitive field and annulus of the survey and has been imaged with the full (u,v) range. Our image of the source, Figure \ref{fig:figf2}(a), has been restored with a beam that is $\sim4$ times larger than normal to highlight the large-scale structure within the source. The source has a LAS of 690 mas and is dominated by the A and B lensed images to the west and east, respectively. The two images are separated by $\sim340$ mas and appear to have weaker components that are mirrored on either side of the lens. These weaker components may be a small portion of a lensed jet that is tangentially stretched. Our observations recover $\sim54\%$ of the WENSS flux suggesting the presence of structures that are fully resolved out even with our shortest baselines. L-Band VLA images of the source seem to suggest that there is indeed a larger-scale emission surrounding the source \citep{ode92}. The measured redshift of the lensing galaxy $z=0.68466\pm0.00004$ \citep{bro93} and that of the lensed object is $z=0.944\pm0.002$ \citep{coh03}.   B0218.0+3542 will be studied in greater detail with the high sensitivity, narrow-field observations of this source at 327 MHz and at 610 MHz by Wucknitz et al. (in preparation).

\subsubsection{B0221.9$+$3417}
The VLBI source we detect within the B0221.9$+$3417 field, Figure \ref{fig:figf1}(b), is offset by $9.94\arcsec$ and at a position angle of $146\arcdeg$ compared to the WENSS position. The source is also detected in field 2, Figure \ref{fig:figf2}(o). The separation between both detections is within 100 mas, after correcting for the larger-scale offset described in \S~\ref{sec:vlbical}, and both have a similar flux density confirming that the detected source is indeed at this position. The position offset also exists when compared against the same source in NVSS, the 365 MHz Texas survey \citep{dou96} and VLSS \citep{coh07}. As only 11\% of the WENSS flux was recovered by the VLBI observation, the position offset hints at a larger component $\sim10\arcsec$ to the north-west of the VLBI source. Furthermore, NVSS lists a fitted source size with a major axis of $16\arcsec$ and the Texas survey categorises the source as a symmetric double with a component separation of $13\pm2\arcsec$ at a position angle of $155\pm11\arcdeg$. These observations are consistent with the VLBI observation if we assume a compact south-eastern component has been detected. The VLBI source has a weaker component 360 mas to the north-west that is directly in line with the WENSS source. We measure a LAS of 360 mas for the VLBI source and an LLS 2.8 kpc at its measured redshift of $0.852\pm0.002$ \citep{wil02}.

\subsubsection{B0221.6$+$3406B}
B0221.6$+$3406B, as shown in Figure \ref{fig:figf1}(c), has a complex morphology. The source has an LAS of 830 mas which corresponds to a LLS of 6.7 kpc at a measured redshift of $z=2.195\pm0.003$ \citep{wil98}. Our VLBI observations have recovered $\sim90\%$ of the WENSS flux suggesting that there is little or no extended structure above what has already been imaged. Based on the integrated flux densities in WENSS and NVSS, the source has a spectral index of $-0.93$. 

\subsubsection{B0223.9$+$3351}
B0223.9$+$3351, as shown in Figure \ref{fig:figf1}(d), appears to be an AGN with a 180 mas jet extension to the north. The source has a LAS of 420 mas which corresponds to a LLS of 3.5 kpc at a measured redshift of $z=1.245\pm0.004$ \citep{wil02}. Our VLBI observations have recovered $\sim50\%$ of the WENSS flux. 

\subsubsection{B0215.4$+$3536}
B0215.4$+$3536 is an ultra-steep spectrum source, a characteristic that is an excellent tracer of galaxies at redshifts $z\geq2$ \citep[e.g.][and references therein]{rot94}. The source has a LAS of $7.1\arcsec$ which corresponds to a LLS of $\geq60$ kpc for a redshift of $z\geq2$. Figure \ref{fig:figf2}(e) shows our 90 cm VLBI image overlaid with an L-Band VLA image \citep{rot94}. The core and peaks within the two lobes align very closely, to within $0.1\sigma$, with the VLA image. A compact component of emission, possibly a jet interaction region, is detected by our observations approximately mid-way between the core and the south-west lobe and appears to align with extended edge of that lobe. Approximately 50\% of the WENSS flux is recovered by our observation, the remaining flux is likely associated with the extended lobes and is resolved out by our observation.

\subsubsection{B0219.2$+$3339}
B0219.2$+$3339 is detected in both field 1 and field 2 and is shown in Figures \ref{fig:figf1}(g) and \ref{fig:figf2}(s), respectively. A significant residual offset of 200 mas exists between these two independent detections even after correcting for the larger-scale offset described in \S~\ref{sec:vlbical}. The phases of this source in the J0226$+$3421 field appeared to be more heavily affected by the ionosphere than those in the B0218$+$357 field and it is believed that the offset may have been introduced by the phase self-calibration process. The source has a LAS of 900 mas which corresponds to a LLS of 6.6 kpc for a measured redshift of $z=0.752\pm0.002$ \citep{wil02}.

\subsubsection{B0214.5$+$3503}
B0215.4$+$3503 has been previously imaged by \citet{rot94} at L-Band with the VLA and is classified as an ultra-steep spectrum source. Contours of the L-Band VLA image are shown overlaid with the 90 cm VLBI observation in Figure \ref{fig:figf2}(h). The source has a LAS of $2.8\arcsec$ which corresponds to a LLS of $\geq24$ kpc if a redshift of $z\geq2$ is assumed. The two components of the VLBI source align very closely, to within $0.2\sigma$, with the VLA image. Approximately 75\% of the WENSS flux is recovered by our observation, the remaining flux is likely associated with the extended structure observed in the VLA image.

\subsubsection{B0223.5$+$3542}
B0223.5$+$3542 is detected in both field 1 and field 2 and is shown in Figures \ref{fig:figf1}(h) and \ref{fig:figf2}(j), respectively. A significant residual offset of 180 mas exists between these two independent detections even after correcting for the larger-scale offset described in \S~\ref{sec:vlbical}. The phases of the source in the J0226$+$3421 field appeared to be more heavily affected by the ionosphere than those in the B0218$+$357 field and it is believed that the offset may have been introduced by the phase self-calibration process. This may have also adversely affected the measure of the integrated flux density which is approximately 60\% greater in the J0226$+$3421 field compared to the B0218$+$357 field. The two components of the source are separated by 830 mas. Approximately 90\% of the WENSS flux is recovered in the B0218$+$357 field observation of this source.

\subsubsection{B0223.8$+$3533}
Observations of this field revealed an unresolved compact source, Figure \ref{fig:figf2}(k), approximately $17\arcsec$ from the position reported by WENSS and over $25\arcsec$ from the more accurate position reported by NVSS, these are by far the greatest offsets observed in all of our detected sources. NVSS also places a limit on the fitted major axis of this source at less than $15\arcsec$, a size that does not encompass our observed source. Furthermore, our measure of integrated flux of 90 mJy is more than twice that measured by WENSS. The size, position and integrated flux density of our VLBI source suggest that it may be an unrelated transient or highly variable source that has been serendipitously detected within the target field. The existence of this source was tested by splitting the VLBI data into four equal length periods and independently imaging each of these. The source was detected in all four data sets with an integrated flux of $92\pm11$ mJy beam$^{-1}$. Furthermore, as would be expected for an unresolved source, scalar averaging of the visibility amplitudes over 30 minute intervals revealed amplitudes that were approximately equal on all baselines.

\subsubsection{B0211.9$+$3452}
We observe an almost $7\arcsec$ offset at a position angle of $13\arcdeg$ between our VLBI detection of this source, Figure \ref{fig:figf2}(l), and the best known position \citep{con98}. NVSS places a limit on the fitted HWHM model of this source at less than $28.6\arcsec$ and also notes the presence of large residual errors which are indicative of a complex source.  As only 30\% of the WENSS flux was recovered by the VLBI observation there is a suggestion that we have detected a compact component of a larger source. This is also supported by the 365 MHz Texas survey which classifies the source as an asymmetric double with a separation of $27\pm1\arcsec$ at a position angle of $25\pm2\arcdeg$.

\section{Discussion}

Our survey results indicate that at least $10\%$ of moderately faint (S$\sim100$ mJy) sources found at 90 cm contain compact components smaller than $\sim0.1$ to $0.3$ arcsec and stronger than $10\%$ of their total flux densities. This is a strict lower limit as the sensitivity of our observation was limited by the primary beam at the edge of the survey fields. None of the surveyed sources that were even slightly resolved by WENSS were detected. Similarly, none of the WENSS sources that were below the sensitivity limits of the VLBI observation were detected either, suggesting that none of these sources had significantly increased in brightness since the WENSS observations were carried out.  

The apparent lack of sources varying above our detection threshold must at least in part be due to resolution effects.  As 90\% of the WENSS sources above the VLBI detection threshold are not detected they must be at least partially resolved at the VLBI resolution and the compact component of the radio emission would only be a fraction of the WENSS flux density.  For the compact component of these sources to vary enough to be detected with VLBI, they must increase in strength by factors of perhaps at least a few (the reciprocal of the ratio of compact flux to WENSS flux) to be detectable with VLBI; the compact component of the flux needs to increase above the VLBI sensitivity limit.  Resolution effects are masking variability in these sources.  As discussed below, the detection of one apparent highly variable source in the VLBI data is rather remarkable.

The interpretation of our detection statistics is complicated, in that the survey has a non-uniform sensitivity over both fields, due to the primary beam response of the VLBA antennas and the fact that we are imaging objects well beyond the half-power points of the primary beam.  In addition, due to time and bandwidth smearing effects, as one images objects further from the phase centre, data on the long baselines is discarded, since the smearing effects make imaging difficult.  A consequence of this is that the angular resolution is also non-uniform across the surveyed fields, with low resolution far from the phase centre.  Not only is the flux limit variable across the field, the brightness temperature sensitivity also varies.

It is possible to estimate the detection statistics of our survey for a uniform flux density and brightness temperature limit by considering sources not too far from the phase centre and for a flux sensitivity between the extremes at the phase centre and field edge.  For example, if a sensitivity limit of 30 mJy beam$^{-1}$ is considered (achieved in the $0.25 - 0.5$ degree annulus of field 1 and in the $0.5 - 1$ degree annulus of field 2, and exceeded in the lower radius annuli in each field), 11 out of 55 possible sources are detected, a detection rate of 20\%, higher than the strict lower limit of 10\% estimated above for all sources at all annuli.

\citet{gar05} performed a similar survey of the NOAO Bootes field at 1.4 GHz, using the NRAO VLBA and 100 m Green Bank Telescope. The survey covered a total of 0.28 deg$^{2}$, one hundredth of the area covered by our survey, and detected a total of 9 sources. The survey achieved sensitivities of $0.074-1.2$ mJy beam$^{-1}$ that enabled the detection of both weak and extended sources, whereas our 90 cm observations detected mainly compact sources or slightly resolved bright sources. Nonetheless, we can estimate the number of detections in this region that could be achieved using the 90 cm survey techniques described in this paper. The 0.28 deg$^2$ NOAO Bootes field contains a total of 13 WENSS sources, 6 of which have integrated flux densities $>30$ mJy. Based on our detection rate of 20\% for such sources we would expect to detect one WENSS source at 90 cm. Assuming a median spectral index of -0.77, only two of the \citet{gar05} sources have integrated flux densities above our 30 mJy beam$^{-1}$ limit at 90 cm, however, one of these is extended and would have a VLBI peak flux density that falls below our limit. Thus the observations of \citet{gar05} are consistent with our 90 cm VLBI results for sources with a peak flux density above 30 mJy beam$^{-1}$.

Estimates of the percentage of sources detected with VLBI gives an estimate of the relative contribution of AGN (that contain compact radio emission and are detectable with VLBI) and starburst galaxies (which contain low brightness temperature radio emission not detectable with VLBI).  Analysis of the ratio of starburst galaxies to AGN as a function of redshift (at high redshifts) can help to determine the initial sources of ionising radiation early in the Universe.  As very little redshift data for our surveyed sources are available, such an analysis is not currently possible with this dataset.  In practice, VLBI data at an additional frequency is also required, to confirm that the compact radio emission attributed to AGN has plausible spectral indices.

The distribution of morphologies in the detected survey sources are typical of AGN.  10/27 sources are unresolved point sources, consistent with core-dominated AGN.  A further 8/27 are clearly resolved into double component sources, consistent with being core-jet AGN or double-lobed radio galaxies.  7/27 sources have complex or extended structures, not obviously clear double components.  Again, these sources may be core-jet AGN or radio galaxies.  The remaining 2/27 sources are the gravitational lens and the quasar at the phase centres of the two fields.

The serendipitous detection of a likely highly variable, very compact source near the target WENSS source B0223.8$+$3533 is intriguing. The total area imaged by this survey represents $\sim0.5\%$ of the area within the $0\arcdeg-2\arcdeg$ annulus and is equivalent to $\sim2.2\%$ of the FWHM of the VLBA primary beam. While it is difficult to place any limits on the real population of variable sources based on this one observation, it does highlight the importance of imaging wide-fields completely, in order to improve our understanding of such sources.

\subsection{Future Prospects}

The observations presented here demonstrate that extremely wide-field surveys can now be piggybacked on current and future VLBI observations at 90 cm. While this survey has mainly concentrated on detecting and imaging sources already detected by other surveys, we find tantalising evidence of a transient or highly variable source. We were fortunate to have found one that appeared in close proximity to one of our target sources but this may not always be the case. This provides a motive to take on a more ambitious survey of the entire field. Such a survey is not beyond the reach of current technology, it would require at most $\sim45$ times more processing compared to the project presented here, in order to image the entire primary beam of the VLBA using a similar faceted approach. While this is not the most efficient means of detecting transients, it will help progress the development of algorithms and techniques needed for next generation, survey-class instruments that operate at wavelengths or sensitivities not matched by current instruments.

The observations presented in this paper were limited by the spectral and temporal resolution of the EVN correlator at the time of the observation. To minimise the effects of bandwidth and time-averaging smearing it was necessary to compromise resolution and image noise. Future technical developments in the capabilities of correlators will allow wide-field, global VLBI studies to be conducted without such restrictions. In particular, software correlators can provide extremely high temporal and spectral resolution, limited only by the time it takes to process the data \citep{del07}. They also allow for some pre-processing to be applied during the correlation process to, for example, mitigate the effects of radio interference or to correlate against multiple phase centres simultaneously.

\subsection{Implications for LOFAR and SKA}

The results of these observations provide important information on the nature and incidence of compact, low-frequency radio sources, with consequences for next generation, low-frequency instruments such as LOFAR and the SKA. LOFAR is currently being deployed across The Netherlands but remote stations are already under construction in neighbouring countries, in particular Germany. Other countries (e.g. UK, France, Sweden, Italy and Poland) are also expected to join this European expansion of LOFAR (E-LOFAR), extending the longest baseline from a few hundred, to a few thousand km. This development will provide LOFAR with sub-arcsecond resolution at its highest observing frequency (the $120-240$ MHz high-band). One concern associated with extending LOFAR to much longer baselines is whether enough cosmic sources will remain unresolved - this characteristic is required in order to ensure there are enough calibrator sources in the sky in order to calibrate the instrument across its full, very wide, field-of-view.  The observations presented here suggest that at least one tenth of all radio sources (at the several tens of mJy level) are likely to exhibit compact VLBI radio structure in the LOFAR high-band. In all likelyhood, an even larger fraction of the E-LOFAR source population will therefore be bright and compact enough to form a grid of calibrator sources across the sky. From our results, we estimate the spatial number density of relatively bright (S$>10$ mJy) and compact (LAS$<200$ mas) sources at 240 MHz to be $\sim3$ deg$^{-2}$. The aggregate total of these compact sources within a beam should serve as a good calibrator for E-LOFAR and enable most of the low-frequency radio sky to be imaged with excellent sub-arcsecond resolution and high dynamic range. Extrapolation to LOFAR's low-band (10$-$80 MHz) is probably very dangerous, but there is every reason to believe that a large number of these sources will remain compact.

In order to assess the relative numbers of starburst galaxies and AGN as a function of redshift, obviously large redshift surveys need to take place for these radio continuum objects.  Such a survey could be conducted using the redshifted HI signal from these galaxies, using the SKA.

\acknowledgements

E.L. acknowledges support from a Swinburne University of Technology Chancellor's Research Scholarship, a CSIRO Postgraduate Student Research Scholarship, ATNF co-supervision, and the hospitality of JIVE where part of this work was carried out. This work was supported by the European Community's Sixth Framework Marie Curie Research Training Network Programme, Contract No. MRTN-CT-2004-505183 ``ANGLES''. This research has made use of the NASA/IPAC Extragalactic Database (NED), which is operated by the Jet Propulsion Laboratory, California Institute of Technology, under contract with the National Aeronautics and Space Administration. The National Radio Astronomy Observatory is a facility of the National Science Foundation operated under cooperative agreement by Associated Universities, Inc. The European VLBI Network is a joint facility of European, Chinese, South African and other radio astronomy institutes funded by their national research councils.

\clearpage

\begin{deluxetable}{lcccccccc}
\tabletypesize{\scriptsize}
\tablecolumns{9}
\tablecaption{Survey Fields, Depths and Source Counts at 324 MHz.}
\tablehead{
   \colhead{Survey}                         &
   \colhead{Annulus}                        &
   \colhead{Maximum}                        &
   \colhead{Response}                       &
   \colhead{$1\sigma$ rms}                  &
   \colhead{Survey}                         &
   \colhead{$S_{P}$\tablenotemark{b}}       &
   \colhead{$N_{WENSS}$}                    &
   \colhead{$\langle N_{VLBI}\rangle$\tablenotemark{c}}  \\
   \colhead{Field\tablenotemark{a}}         &
   \colhead{Range}                          &
   \colhead{$(u, v)$ Range}                 &
   \colhead{R\tablenotemark{b}}             &
   \colhead{Noise\tablenotemark{b}}         &
   \colhead{Resolution}                     &
   \colhead{}                               &
   \colhead{}                               &
   \colhead{}                               \\
   \colhead{}                               &
   \colhead{(arcdeg)}                       &
   \colhead{(M$\lambda$)}                   &
   \colhead{}                               &
   \colhead{(mJy beam$^{-1}$)}              &
   \colhead{(mas, mas)}                     &
   \colhead{(mJy beam$^{-1}$)}
}
\startdata
1 & $0.00-0.25$  & 2.50   & 0.94  & 3.7   & $40\times20$    & $>24$  &  4    &     3     \\
1 & $0.25-0.50$  & 1.00   & 0.89  & 4.8   & $140\times130$  & $>32$  &  15   &     10    \\
1 & $0.50-1.00$  & 0.75   & 0.65  & 6.8   & $180\times170$  & $>63$  &  55   &     27    \\
1 & $1.00-1.50$  & 0.75   & 0.36  & 8.9   & $230\times220$  & $>147$ &  91   &     28    \\
1 & $1.50-2.00$  & 0.50   & 0.17  & 11.0  & $350\times290$  & $>378$ &  128  &     16    \\
1 & $2.00-3.00$  & 0.50   & 0.06  & 15.1  & $360\times290$  & $>1603$&  411  &     11    \\ \hline
2 & $0.00-0.25$  & 2.50   & 0.94  & 1.9   & $90\times80$    & $>12$  &  2    &     2     \\
2 & $0.25-0.50$  & 1.00   & 0.89  & 2.3   & $140\times130$  & $>16$  &  15   &     15    \\
2 & $0.50-1.00$  & 0.75   & 0.65  & 3.2   & $180\times170$  & $>30$  &  50   &     44    \\
2 & $1.00-1.50$  & 0.75   & 0.36  & 4.1   & $230\times220$  & $>68$  &  97   &     52    \\
2 & $1.50-2.00$  & 0.50   & 0.17  & 5.1   & $350\times290$  & $>174$ &  128  &     41    \\
2 & $2.00-3.00$  & 0.50   & 0.06  & 6.9   & $360\times290$  & $>749$ &  375  &     23    
\enddata
\tablenotetext{a}{Field 1 is centred about J0226$+$3421 and field 2 is centred about B0218$+$357.}
\tablenotetext{b}{The estimated worst case values at the edge of the annulus.}
\tablenotetext{c}{The number of WENSS sources that, if unresolved, have a peak flux that would fall above the estimated VLBI sensitivity limit.}
\label{tab:tabfields}
\end{deluxetable}

\begin{deluxetable}{lccccccccc}
\tabletypesize{\scriptsize}
\tablecolumns{10}
\tablecaption{Source and image characteristics and astrometric errors.}
\tablehead{
   \colhead{WENSS}               &
   \multicolumn{2}{c}{Source Location}            &
   \multicolumn{3}{c}{Image Characteristics}      &
   \multicolumn{4}{c}{Position Comparison}  \\
   \colhead{Source}              &
   \colhead{Field\tablenotemark{a}}               &
   \colhead{$d_{PC}$\tablenotemark{b}}            &
   \colhead{Beam Size}           &
   \colhead{Beam P.A.}           &
   \colhead{$1\sigma$ Noise\tablenotemark{c}}     &
   \colhead{Ref.\tablenotemark{d}}                &
   \colhead{$d_{E}$\tablenotemark{e}}             &
   \colhead{$\theta_{E}$\tablenotemark{f}}        &
   \colhead{$d_{\sigma}$\tablenotemark{g}}        \\
   \colhead{}                    &
   \colhead{}                    &
   \colhead{(arcdeg)}            &
   \colhead{(mas,mas)}           &
   \colhead{(degrees)}           &
   \colhead{(mJy beam$^{-1}$)}   &
   \colhead{}                    &
   \colhead{(arcsec)}            &
   \colhead{(degrees)}           &
   \colhead{($\sigma_{p}$)}
}
\startdata
3C84      & \nodata & \nodata & $31\times15$   & $-17$  & 6.1 & BE02     & 0.002  & 228  & 0.7   \\
B0223.1+3408  &  1  &  0.00   & $37\times19$   & $-13$  & 1.8 & BE02     & 0.002  & 270  & 0.7   \\
\nodata       &  2  &  1.89   & $360\times285$ & $-42$  & 5.8 & BE02     & 0.042  & 125  & 1.4   \\
B0218.0+3542  &  2  &  0.00   & $64\times43$   & $52$   & 1.0 & PA92     & 0.024  & 75   & 0.9   \\
B0219.1+3533  &  2  &  0.27   & $88\times84$   & $80$   & 1.5 & CO98     & 0.173  & 168  & 0.1   \\
B0221.9+3417  &  1  &  0.29   & $145\times132$ & $38$   & 4.4 & DO96     & 9.938  & 34   & 13.2  \\
\nodata       &  2  &  1.63   & $361\times285$ & $-42$  & 4.1 & DO96     & 10.003 & 34   & 13.0  \\
B0221.6+3406B &  1  &  0.31   & $145\times132$ & $37$   & 4.1 & CO98     & 0.613  & 160  & 0.6   \\
B0223.9+3351  &  1  &  0.32   & $145\times132$ & $38$   & 4.3 & CO98     & 1.432  & 55   & 1.5   \\
B0221.6+3406A &  1  &  0.33   & $145\times132$ & $37$   & 4.3 & CO98     & 1.178  & 42   & 0.8   \\
B0216.6+3523  &  2  &  0.42   & $142\times133$ & $31$   & 2.2 & CO98     & 0.437  & 188  & 0.4   \\
B0220.1+3532  &  2  &  0.46   & $142\times134$ & $30$   & 2.0 & CO98     & 0.215  & 313  & 0.2   \\
B0215.4+3536  &  2  &  0.54   & $180\times170$ & $23$   & 2.5 & R\"{O}94 & 0.103  & 56   & 0.1   \\
B0223.2+3441  &  1  &  0.55   & $180\times168$ & $38$   & 5.0 & CO98     & 1.382  & 317  & 1.1   \\
B0220.7+3551  &  2  &  0.56   & $180\times170$ & $23$   & 2.4 & CO98     & 0.231  & 148  & 0.2   \\
B0215.8+3626  &  2  &  0.87   & $227\times217$ & $-38$  & 3.0 & CO98     & 0.226  & 354  & 0.2   \\
B0219.2+3339  &  1  &  0.93   & $224\times218$ & $-62$  & 6.3 & CO98     & 0.572  & 46   & 0.6   \\
\nodata       &  2  &  2.06   & $359\times284$ & $-43$  & 4.2 & CO98     & 0.374  & 42   & 0.4   \\
B0214.5+3503  &  2  &  0.96   & $227\times217$ & $-36$  & 3.9 & R\"{O}94 & 0.162  & 49   & 0.2   \\
B0215.4+3636  &  2  &  1.04   & $228\times217$ & $-36$  & 3.3 & CO98     & 0.579  & 114  & 0.6   \\
B0223.5+3542  &  2  &  1.12   & $228\times217$ & $-37$  & 5.3 & DO96     & 0.833  & 295  & 1.8   \\
\nodata       &  1  &  1.57   & $357\times287$ & $-44$  & 9.1 & DO96     & 0.783  & 308  & 1.6   \\
B0223.8+3533  &  2  &  1.19   & $228\times217$ & $-35$  & 3.1 & CO98     & 25.146 & 344  & 11.0  \\
B0211.9+3452  &  2  &  1.50   & $227\times217$ & $-37$  & 3.6 & CO98     & 6.936  & 344  & 10.6  \\
B0214.3+3700  &  2  &  1.51   & $352\times286$ & $-40$  & 4.6 & CO98     & 2.544  & 271  & 2.6   \\
B0215.1+3710  &  2  &  1.58   & $353\times287$ & $-41$  & 4.1 & CO98     & 0.368  & 322  & 0.4   \\
B0226.5+3618  &  2  &  1.81   & $360\times288$ & $-40$  & 3.8 & CO98     & 1.362  & 192  & 1.4   \\
B0225.1+3659  &  2  &  1.92   & $358\times288$ & $-40$  & 5.3 & CO98     & 0.256  & 143  & 0.3   
\enddata
\tablenotetext{a}{Field 1 is centred about J0226+3421 and field 2 is centred about B0218$+$357.}
\tablenotetext{b}{Distance of the source from the phase centre of the observed field.}
\tablenotetext{c}{The $1\sigma$ residual noise after the source model has been subtracted from the image.}
\tablenotetext{d}{References for positions.~
  BE02 = 2.3/8.3 GHz VLBA \citep{bea02};
  CO98 = 1.4 GHz VLA \citep{con98};
  DO96 = 365 MHz Texas interferometer \citep{dou96};
  PA92 = 8.4 GHz VLA \citep{pat92};
  R\"{O}94 = 1.4 GHz VLA \citep{rot94}
}
\tablenotetext{e}{The measured offset of the VLBI source peak from the position listed in reference.}
\tablenotetext{f}{The position angle of the VLBI source peak in relation to the position listed in reference.}
\tablenotetext{g}{The measured offset of the VLBI source peak in relation to the position listed in reference in terms of the $1\sigma$ astrometric precision $\sigma_{p}$.}
\label{tab:tabastrometry}
\end{deluxetable}

\begin{deluxetable}{llllccc}
\tabletypesize{\scriptsize}
\tablecolumns{7}
\tablecaption{Astrometry and Photometry at 324 MHz for field 1 targets.}
\tablehead{
   \colhead{Annulus}             &
   \colhead{WENSS Source}        &
   \colhead{R.A.}                &
   \colhead{Decl.}               &
   \colhead{$S_{P}$}             &
   \colhead{$S_{I}$}             &
   \colhead{$\alpha$}            \\
   \colhead{}                    &
   \colhead{}                    &
   \colhead{(J2000.0)}           &
   \colhead{(J2000.0)}           &
   \colhead{(mJy beam$^{-1}$)}   &
   \colhead{(mJy)}               &
   \colhead{}
}
\startdata
$0\arcdeg-0.25\arcdeg$ & B0223.1+3408 & 02 26 10.15   & +34 21 30.7   & 3793 & 3678 & $-0.16$   \\
& \nodata      & 02 26 10.3332 & +34 21 30.286 & 375  & 2880 & \nodata \\
& B0222.2+3406 & 02 25 14.12   & +34 20 22.4   & 32   & 33   & $-1.09$   \\
& B0222.3+3414 & 02 25 19.18   & +34 27 31.7   & 41   & 50   & \nodata \\
\tableline
$0.25\arcdeg-0.5\arcdeg$ & B0221.9+3417  & 02 24 56.18  & +34 30 35.8   & 1657 & 1638 & $-0.91$   \\
& \nodata       & 02 24 56.620 & +34 30 27.47  & 111  & 186  & \nodata \\
& B0224.1+3354  & 02 27 09.72  & +34 08 11.4   & 35   & 17   & $-0.94$   \\
& B0221.6+3406B & 02 24 41.08  & +34 19 43.6   & 459  & 433  & $-0.93$   \\
& \nodata       & 02 24 41.106 & +34 19 41.88  & 78   & 382  & \nodata \\
& B0221.6+3406  & 02 24 39.96  & +34 20 03.1   & 459  & 543  & \nodata \\
& B0223.9+3351  & 02 26 55.50  & +34 04 55.7   & 532  & 525  & $-1.02$   \\
& \nodata       & 02 26 55.564 & +34 04 55.57  & 114  & 269  & \nodata \\
& B0221.6+3406A & 02 24 34.48  & +34 21 39.5   & 104  & 109  & $-1.11$   \\
& \nodata       & 02 24 34.514 & +34 21 35.03  & 97   & 84   & \nodata \\
& B0224.8+3406  & 02 27 54.18  & +34 19 39.5   & 83   & 73   & $-0.88$   \\
& B0221.3+3405  & 02 24 20.16  & +34 19 17.5   & 195  & 193  & $-0.84$   \\
& B0224.9+3419  & 02 27 59.75  & +34 32 46.5   & 50   & 33   & $0.15$    \\
& B0224.2+3430  & 02 27 16.40  & +34 43 42.9   & 129  & 128  & $-0.96$   \\
\tableline
$0.5\arcdeg-1\arcdeg$ & B0220.7+3401  & 02 23 47.53  & +34 15 08.3   & 403  & 421 & $-0.80$   \\
& B0225.5+3419A & 02 28 29.29  & +34 34 43.2   & 89   & 75  & $-0.33$   \\
& B0225.5+3419  & 02 28 34.05  & +34 33 09.5   & 504  & 568 & $-1.06$   \\
& B0225.5+3419B & 02 28 34.63  & +34 32 58.5   & 504  & 492 & $-0.97$   \\
& B0223.2+3441  & 02 26 17.12  & +34 54 36.5   & 88   & 85  & $-0.71$   \\
& \nodata       & 02 26 17.184 & +34 54 36.40  & 126  & 119 & \nodata \\
& B0224.8+3434  & 02 27 51.46  & +34 47 36.1   & 100  & 99  & $-0.65$   \\
& B0223.5+3443  & 02 26 36.80  & +34 56 42.9   & 38   & 33  & $-0.39$   \\
& B0225.5+3347  & 02 28 34.76  & +34 01 00.2   & 148  & 139 & $-0.71$   \\
& B0224.7+3442  & 02 27 43.89  & +34 56 23.1   & 100  & 85  & $-0.90$   \\
& B0221.0+3440  & 02 24 03.11  & +34 54 26.6   & 107  & 102 & $-0.41$   \\
& B0225.7+3340  & 02 28 45.59  & +33 53 31.1   & 95   & 84  & $-0.66$   \\
& B0220.4+3433  & 02 23 25.14  & +34 47 20.1   & 134  & 145 & \nodata \\
& B0219.6+3404  & 02 22 40.67  & +34 18 10.7   & 54   & 41  & $-0.87$   \\
& B0225.8+3438A & 02 28 48.50  & +34 51 32.8   & 87   & 86  & $-0.85$   \\
& B0219.6+3415  & 02 22 38.27  & +34 29 29.1   & 188  & 174 & $-0.81$   \\
& B0225.8+3438  & 02 28 52.37  & +34 52 17.3   & 87   & 161 & \nodata   \\
& B0225.8+3438B & 02 28 56.96  & +34 53 04.5   & 81   & 74  & $-0.92$   \\
& B0219.8+3343A & 02 22 48.79  & +33 57 18.9   & 638  & 638 & $-0.83$   \\
& B0219.8+3343  & 02 22 48.85  & +33 57 11.3   & 638  & 688 & $-0.88$   \\
& B0221.2+3325  & 02 24 11.73  & +33 39 18.1   & 185  & 178 & $-0.78$   \\
& B0219.8+3343B & 02 22 49.91  & +33 55 04.3   & 49   & 49  & $-0.84$   \\
& B0218.8+3402  & 02 21 50.01  & +34 16 19.1   & 359  & 347 & $-0.88$   \\
& B0223.5+3501  & 02 26 32.36  & +35 15 21.8   & 122  & 102 & $-0.81$   \\
& B0218.9+3353  & 02 21 56.58  & +34 07 17.5   & 127  & 122 & $-0.92$   \\
& B0227.5+3414  & 02 30 35.36  & +34 27 45.3   & 74   & 67  & $-0.86$   \\
& B0219.2+3339  & 02 22 16.75  & +33 53 27.3   & 516  & 499 & $-0.67$   \\
& \nodata       & 02 22 16.812 & +33 53 26.90  & 251  & 417 & \nodata \\
& B0227.0+3335  & 02 30 01.40  & +33 48 48.3   & 142  & 138 & \nodata \\
\tableline
$1\arcdeg-1.5\arcdeg$ & B0218.6+3342  & 02 21 40.00 & +33 56 36.1 & 195  & 204  & $-0.84$   \\
& B0227.6+3343  & 02 30 41.94 & +33 56 23.3 & 68   & 59   & $-0.35$   \\
& B0225.4+3312  & 02 28 25.15 & +33 26 14.0 & 128  & 111  & $-0.63$   \\
& B0226.5+3321  & 02 29 31.62 & +33 34 23.8 & 215  & 207  & $-0.79$   \\
& B0218.1+3419  & 02 21 09.69 & +34 33 37.8 & 252  & 252  & $-0.54$   \\
& B0220.5+3310  & 02 23 33.90 & +33 24 35.5 & 179  & 175  & $-0.80$   \\
& B0227.7+3443  & 02 30 44.68 & +34 57 14.6 & 98   & 96   & $-0.53$   \\
& B0227.3+3325  & 02 30 19.97 & +33 38 53.8 & 76   & 68   & $-0.80$   \\
& B0219.1+3323  & 02 22 06.57 & +33 36 49.0 & 123  & 112  & $-0.88$   \\
& B0224.6+3515  & 02 27 38.33 & +35 28 43.8 & 164  & 147  & $-0.82$   \\
& B0227.7+3324  & 02 30 42.60 & +33 37 59.8 & 1029 & 996  & $-0.77$   \\
& B0226.1+3509  & 02 29 11.30 & +35 22 22.6 & 160  & 178  & \nodata   \\
& B0217.9+3442  & 02 20 56.23 & +34 56 36.2 & 94   & 104  & $-0.72$   \\
& B0225.0+3258  & 02 28 03.91 & +33 11 27.5 & 399  & 388  & $-0.96$   \\
& B0228.9+3429  & 02 31 59.53 & +34 42 26.7 & 2361 & 2556 & $-1.02$   \\
& B0227.9+3322  & 02 30 56.80 & +33 35 16.5 & 477  & 433  & $-0.77$   \\
& B0225.5+3255  & 02 28 30.73 & +33 09 16.4 & 161  & 131  & $-0.68$   \\
& B0229.4+3410  & 02 32 28.67 & +34 24 03.0 & 9334 & 9459 & $-0.88$   \\
& B0217.8+3452B & 02 20 49.19 & +35 06 28.3 & 107  & 98   & $-1.30$   \\ 
& B0217.8+3452  & 02 20 48.92 & +35 06 25.2 & 107  & 109  & $-1.37$   \\
& B0216.6+3409  & 02 19 37.82 & +34 23 10.5 & 555  & 515  & $-1.06$   \\
& B0229.0+3332  & 02 32 05.82 & +33 45 22.2 & 321  & 305  & $-1.24$   \\
& B0220.8+3527  & 02 23 50.17 & +35 41 31.7 & 398  & 393  & $-0.83$   \\
& B0216.2+3405B & 02 19 18.40 & +34 19 49.5 & 273  & 338  & \nodata   \\
& B0216.2+3405  & 02 19 15.49 & +34 19 42.1 & 412  & 802  & \nodata   \\
& B0216.8+3334  & 02 19 48.77 & +33 48 16.0 & 1352 & 1357 & \nodata   \\
& B0229.3+3447  & 02 32 21.61 & +35 01 06.8 & 352  & 381  & $-1.30$   \\
& B0216.2+3405A & 02 19 13.44 & +34 19 37.2 & 412  & 463  & \nodata   \\
\tableline
$1.5\arcdeg-2\arcdeg$ & B0230.3+3429  & 02 33 20.42  & +34 42 54.1   & 170  & 172  & $0.06$    \\
& B0229.3+3318  & 02 32 21.28  & +33 32 08.4   & 168  & 150  & $-0.77$   \\
& B0223.5+3542  & 02 26 36.14  & +35 55 46.2   & 1758 & 1730 & $-0.59$   \\
& \nodata       & 02 26 36.114 & +35 55 45.44  & 675  & 1491 & \nodata   \\
& B0226.2+3534  & 02 29 20.00  & +35 47 40.2   & 167  & 163  & $-0.80$   \\
& B0220.1+3238  & 02 23 09.03  & +32 51 58.0   & 206  & 257  & \nodata   \\
& B0220.1+3238A & 02 23 08.00  & +32 51 43.3   & 206  & 191  & $-0.77$   \\
& B0216.2+3318  & 02 19 11.54  & +33 31 49.9   & 302  & 284  & $-0.53$   \\
& B0229.4+3517  & 02 32 29.32  & +35 30 53.6   & 2149 & 2359 & $-0.86$   \\
& B0231.2+3500  & 02 34 15.22  & +35 13 53.0   & 294  & 300  & $-0.95$   \\
& B0219.6+3547  & 02 22 37.70  & +36 01 01.1   & 487  & 515  & $-0.93$   \\
& B0218.0+3542  & 02 21 05.40  & +35 56 13.0   & 2534 & 2460 & $-0.25$   \\
& B0232.4+3405  & 02 35 26.28  & +34 18 30.3   & 391  & 384  & $-0.55$   \\
& B0214.7+3321  & 02 17 43.31  & +33 35 03.0   & 683  & 657  & $-0.85$   \\
& B0223.0+3603  & 02 26 04.82  & +36 17 10.0   & 337  & 355  & $-1.11$   \\
& B0223.0+3603A & 02 26 04.43  & +36 17 19.8   & 337  & 308  & $-1.02$   \\
& B0214.5+3503  & 02 17 35.04  & +35 17 22.6   & 1146 & 1092 & \nodata   \\
\tableline
$2\arcdeg-3\arcdeg$ & B0217.8+3227  & 02 20 48.04 & +32 41 07.0 & 1415 & 1393 & $-0.28$   \\
& B0230.1+3243  & 02 33 12.09 & +32 56 48.5 & 594  & 606  & $-0.98$   \\
& B0230.1+3243B & 02 33 12.37 & +32 56 51.2 & 594  & 572  & $-0.94$   \\
& B0215.4+3536  & 02 18 28.69 & +35 49 58.6 & 1445 & 1401 & $-0.89$   \\
& B0227.2+3206B & 02 30 16.48 & +32 20 27.1 & 642  & 685  & \nodata   \\
& B0227.2+3206  & 02 30 14.54 & +32 19 48.0 & 713  & 1473 & \nodata   \\
& B0227.2+3206A & 02 30 12.82 & +32 19 12.7 & 713  & 788  & $-1.06$   \\
& B0233.2+3458  & 02 36 19.28 & +35 11 15.2 & 871  & 843  & $-0.97$   \\
& B0211.9+3452  & 02 14 58.14 & +35 06 39.5 & 3011 & 3084 & $-0.83$   \\
& B0224.9+3650  & 02 28 01.68 & +37 03 36.3 & 1409 & 1455 & $-0.90$   \\
& B0225.1+3659  & 02 28 13.67 & +37 12 57.1 & 1566 & 1519 & $-1.09$   
\enddata
\tablecomments{Units of right ascension are hours, minutes and seconds, and units of declination are degree, arcminutes and arcseconds. The first data entry for each target refers to the WENSS target source, and the second entry, when present, refers to the VLBI detection. The spectral index $\alpha$ is defined as $S_{\nu}\propto\nu^{\alpha}$ and is estimated from the WENSS and NVSS integrated flux densities where available.}
\label{tab:tabf1}
\end{deluxetable}

\begin{deluxetable}{llllccc}
\tabletypesize{\scriptsize}
\tablecolumns{7}
\tablecaption{Astrometry and Photometry at 324 MHz for field 2 targets.}
\tablehead{
   \colhead{Annulus}             &
   \colhead{WENSS Source}        &
   \colhead{R.A.}                &
   \colhead{Decl.}               &
   \colhead{$S_{P}$}             &
   \colhead{$S_{I}$}             &
   \colhead{$\alpha$}            \\
   \colhead{}                    &
   \colhead{}                    &
   \colhead{(J2000.0)}           &
   \colhead{(J2000.0)}           &
   \colhead{(mJy beam$^{-1}$)}   &
   \colhead{(mJy)}               &
   \colhead{}
}
\startdata
$0\arcdeg-0.25\arcdeg$ & B0218.0+3542 & 02 21 05.40   & +35 56 13.0   & 2534 & 2460 & $-0.25$   \\
& \nodata      & 02 21 05.4720 & +35 56 13.716 & 90   & 1320 & \nodata \\
& B0218.8+3545 & 02 21 55.21   & +35 59 20.4   & 20   & 22   & $-0.90$ \\
\tableline
$0.25\arcdeg-0.5\arcdeg$ & B0219.1+3533 & 02 22 10.16  & +35 46 48.3   & 89  & 87  & $-1.08$   \\
& \nodata      & 02 22 10.123 & +35 46 51.67  & 48  & 85  & \nodata   \\
& B0219.6+3547 & 02 22 37.70  & +36 01 01.1   & 487 & 515 & $-0.93$   \\
& B0217.3+3600 & 02 20 23.66  & +36 14 36.8   & 174 & 178 & $-0.96$   \\
& B0219.6+3533 & 02 22 37.61  & +35 47 30.9   & 20  & 19  & $-1.02$   \\
& B0216.9+3526 & 02 19 57.00  & +35 40 41.9   & 31  & 32  & $-1.33$   \\
& B0218.8+3604 & 02 21 55.81  & +36 17 49.5   & 32  & 29  & $-1.18$   \\
& B0218.5+3518 & 02 21 34.18  & +35 32 12.5   & 58  & 44  & $-0.78$   \\
& B0216.6+3523 & 02 19 42.13  & +35 37 43.9   & 48  & 59  & $-0.24$   \\
& \nodata      & 02 19 42.305 & +35 37 44.33  & 41  & 84  & \nodata   \\
& B0216.2+3555 & 02 19 18.61  & +36 08 59.4   & 23  & 21  & \nodata   \\
& B0220.0+3552 & 02 23 06.20  & +36 06 06.5   & 22  & 36  & \nodata   \\
& B0220.1+3532 & 02 23 10.75  & +35 45 50.0   & 80  & 76  & $-0.75$   \\
& \nodata      & 02 23 10.747 & +35 45 46.65  & 53  & 97  & \nodata   \\
& B0215.9+3553 & 02 19 00.18  & +36 07 34.5   & 22  & 14  & $-0.12$   \\
& B0219.8+3524 & 02 22 52.98  & +35 38 26.2   & 29  & 29  & $-0.64$   \\
& B0220.3+3537 & 02 23 23.98  & +35 51 05.8   & 35  & 33  & $-0.50$   \\
& B0217.4+3610 & 02 20 26.67  & +36 24 39.8   & 22  & 29  & $-0.88$   \\
\tableline
$0.5\arcdeg-1\arcdeg$ & B0220.4+3531  & 02 23 27.04  & +35 45 29.4   & 25   & 23   & $-0.32$   \\
& B0216.7+3515A & 02 19 44.91  & +35 30 18.8   & 18   & 18   & \nodata   \\
& B0216.7+3515  & 02 19 47.22  & +35 29 00.0   & 48   & 60   & \nodata   \\
& B0216.7+3515B & 02 19 47.92  & +35 28 32.8   & 48   & 41   & $-1.02$   \\
& B0215.4+3536  & 02 18 28.69  & +35 49 58.6   & 1445 & 1401 & $-0.89$   \\
& \nodata       & 02 18 28.996 & +35 50 01.84  & 93   & 670  & \nodata   \\
& B0220.5+3554  & 02 23 36.81  & +36 08 34.1   & 28   & 24   & $-0.83$   \\
& B0220.7+3551  & 02 23 46.60  & +36 05 01.3   & 176  & 173  & $-0.97$   \\
& \nodata       & 02 23 46.790 & +36 05 03.20  & 104  & 149  & \nodata   \\
& B0219.9+3514  & 02 22 56.36  & +35 28 05.4   & 26   & 23   & $-0.73$   \\
& B0220.7+3558A & 02 23 47.14  & +36 11 34.5   & 85   & 89   & $-0.79$   \\
& B0220.7+3558  & 02 23 47.37  & +36 12 17.3   & 85   & 137  & \nodata   \\
& B0220.8+3527  & 02 23 50.17  & +35 41 31.7   & 398  & 393  & $-0.83$   \\
& B0220.7+3558B & 02 23 47.92  & +36 13 49.8   & 45   & 47   & $-0.93$   \\
& B0221.1+3548  & 02 24 08.54  & +36 01 55.2   & 40   & 34   & $-1.06$   \\
& B0221.1+3551  & 02 24 13.51  & +36 04 51.9   & 73   & 67   & $-0.86$   \\
& B0220.5+3609  & 02 23 32.65  & +36 22 47.7   & 22   & 22   & \nodata   \\
& B0215.2+3522  & 02 18 13.64  & +35 36 42.5   & 42   & 39   & $-0.80$   \\
& B0215.4+3516  & 02 18 29.78  & +35 30 27.7   & 110  & 101  & $-1.02$   \\
& B0219.1+3503  & 02 22 07.70  & +35 17 26.9   & 73   & 66   & $-0.95$   \\
& B0221.4+3535A & 02 24 26.91  & +35 48 50.9   & 26   & 19   & $-0.94$   \\
& B0217.8+3500  & 02 20 50.30  & +35 14 26.9   & 52   & 38   & $-0.41$   \\
& B0221.4+3535  & 02 24 29.69  & +35 49 01.3   & 26   & 38   & \nodata   \\
& B0219.2+3622  & 02 22 18.62  & +36 35 49.7   & 61   & 89   & \nodata   \\
& B0216.5+3504  & 02 19 34.57  & +35 18 02.7   & 29   & 25   & $-1.08$   \\
& B0221.4+3535B & 02 24 32.69  & +35 49 11.6   & 26   & 18   & $-0.72$   \\
& B0216.7+3502  & 02 19 44.76  & +35 16 08.4   & 26   & 31   & $-1.06$   \\
& B0215.8+3616  & 02 18 53.75  & +36 30 35.9   & 126  & 119  & $-0.81$   \\
& B0214.4+3534  & 02 17 29.52  & +35 47 56.3   & 105  & 111  & $-0.74$   \\
& B0217.4+3629  & 02 20 30.75  & +36 42 48.3   & 58   & 48   & $-0.52$   \\
& B0219.6+3625  & 02 22 43.74  & +36 39 02.3   & 34   & 24   & $-0.88$   \\
& B0220.5+3618  & 02 23 38.39  & +36 32 08.0   & 30   & 21   & $-0.19$   \\
& B0213.9+3543  & 02 16 59.68  & +35 57 21.6   & 162  & 179  & $-0.72$   \\
& B0218.5+3632  & 02 21 32.74  & +36 45 45.5   & 54   & 63   & $-0.75$   \\
& B0217.8+3452B & 02 20 49.19  & +35 06 28.3   & 107  & 98   & \nodata   \\
& B0217.8+3452  & 02 20 48.92  & +35 06 25.2   & 107  & 109  & \nodata   \\
& B0218.9+3452  & 02 21 57.11  & +35 06 09.7   & 37   & 30   & $-0.6$    \\
& B0215.8+3626  & 02 18 49.99  & +36 40 41.6   & 156  & 144  & $-0.07$   \\
& \nodata       & 02 18 50.038 & +36 40 42.58  & 101  & 126  & \nodata   \\
& B0222.4+3543  & 02 25 28.81  & +35 57 10.2   & 108  & 89   & $-0.68$   \\
& B0213.6+3540  & 02 16 37.63  & +35 54 21.1   & 224  & 234  & $-1.03$   \\
& B0222.6+3554  & 02 25 40.74  & +36 08 19.9   & 48   & 74   & \nodata   \\
& B0222.6+3554B & 02 25 42.54  & +36 07 28.8   & 48   & 45   & $-0.54$   \\
& B0221.4+3622  & 02 24 31.52  & +36 36 02.1   & 33   & 51   & $-0.60$   \\
& B0214.5+3503  & 02 17 35.04  & +35 17 22.6   & 1146 & 1092 & $-0.90$   \\
& \nodata       & 02 17 34.989 & +35 17 21.49  & 187  & 808  & \nodata   \\
& B0214.1+3507  & 02 17 11.78  & +35 21 05.3   & 81   & 81   & $-0.73$   \\
& B0217.9+3442  & 02 20 56.23  & +34 56 36.2   & 94   & 104  & $-0.72$   \\
\tableline
$1\arcdeg-1.5\arcdeg$ & B0222.4+3613A & 02 25 29.65  & +36 25 50.9   & 60   & 91   & \nodata   \\
& B0222.4+3613  & 02 25 30.21  & +36 26 34.7   & 99   & 185  & \nodata   \\
& B0222.4+3613B & 02 25 30.65  & +36 27 08.9   & 99   & 93   & $-0.77$   \\
& B0213.1+3558  & 02 16 09.27  & +36 12 34.7   & 44   & 37   & $-0.67$   \\
& B0223.1+3551  & 02 26 10.41  & +36 04 41.9   & 33   & 29   & \nodata   \\
& B0215.4+3636  & 02 18 30.87  & +36 50 27.2   & 339  & 321  & $-0.55$   \\
& \nodata       & 02 18 30.913 & +36 50 27.45  & 127  & 136  & \nodata   \\
& B0212.9+3535  & 02 15 55.24  & +35 49 12.0   & 61   & 64   & $-0.68$   \\
& B0222.4+3617  & 02 25 29.56  & +36 31 03.3   & 64   & 57   & $-0.85$   \\
& B0217.3+3645  & 02 20 20.42  & +36 59 16.7   & 367  & 2686 & \nodata   \\
& B0223.0+3603A & 02 26 04.43  & +36 17 19.8   & 337  & 308  & $-1.02$   \\
& B0223.0+3603  & 02 26 04.82  & +36 17 10.0   & 337  & 355  & $-1.11$   \\
& B0213.9+3502  & 02 16 57.40  & +35 16 33.5   & 61   & 64   & $-0.77$   \\
& B0223.0+3603B & 02 26 08.11  & +36 15 47.6   & 50   & 47   & \nodata   \\
& B0212.8+3553  & 02 15 48.66  & +36 06 56.9   & 83   & 79   & $-0.80$   \\
& B0212.9+3603  & 02 15 57.29  & +36 16 55.4   & 93   & 91   & $-0.82$   \\
& B0222.7+3618  & 02 25 45.12  & +36 31 32.5   & 57   & 57   & $-0.81$   \\
& B0223.5+3542  & 02 26 36.14  & +35 55 46.2   & 1758 & 1730 & $-0.59$   \\
& \nodata       & 02 26 36.104 & +35 55 45.53  & 518  & 954  & \nodata   \\
& B0217.0+3648B & 02 20 08.92  & +37 02 21.4   & 42   & 54   & $-1.11$   \\
& B0217.0+3648  & 02 20 06.68  & +37 02 41.2   & 45   & 95   & \nodata   \\
& B0217.0+3648A & 02 20 04.01  & +37 03 08.9   & 45   & 40   & $-0.69$   \\
& B0217.4+3434  & 02 20 29.93  & +34 48 25.1   & 62   & 57   & $-1.08$   \\
& B0213.0+3614B & 02 16 02.26  & +36 27 54.2   & 98   & 94   & $-0.81$   \\
& B0213.0+3614  & 02 16 01.21  & +36 27 59.1   & 98   & 117  & \nodata   \\
& B0222.6+3623  & 02 25 44.67  & +36 36 36.7   & 171  & 158  & $-0.55$   \\
& B0223.7+3600  & 02 26 46.21  & +36 13 50.2   & 45   & 36   & $-0.53$   \\
& B0223.8+3533  & 02 26 54.06  & +35 47 03.2   & 48   & 45   & $-0.89$   \\
& \nodata       & 02 26 53.435 & +35 46 48.08  & 86   & 90   & \nodata    \\
& B0221.0+3440  & 02 24 03.11  & +34 54 26.6   & 107  & 102  & $-0.41$   \\
& B0223.6+3607  & 02 26 43.34  & +36 20 55.7   & 109  & 100  & $-0.57$   \\
& B0212.2+3604  & 02 15 18.14  & +36 17 57.2   & 1278 & 1303 & $-0.96$   \\
& B0220.4+3433  & 02 23 25.14  & +34 47 20.1   & 134  & 145  & $-0.98$   \\
& B0224.3+3544  & 02 27 23.05  & +35 57 42.5   & 72   & 75   & $-0.67$   \\
& B0216.3+3656  & 02 19 21.87  & +37 10 05.7   & 106  & 98   & $-0.77$   \\
& B0223.5+3501  & 02 26 32.36  & +35 15 21.8   & 122  & 102  & $-0.81$   \\
& B0216.6+3659  & 02 19 40.76  & +37 13 05.2   & 82   & 89   & $-0.61$   \\
& B0218.1+3419  & 02 21 09.69  & +34 33 37.8   & 252  & 252  & $-0.54$   \\
& B0224.7+3558  & 02 27 48.73  & +36 11 25.1   & 61   & 47   & $-0.02$   \\
& B0211.2+3550  & 02 14 14.41  & +36 04 40.5   & 182  & 186  & $-0.84$   \\
& B0212.4+3454  & 02 15 24.71  & +35 08 50.6   & 78   & 66   & $-0.86$   \\
& B0224.6+3515  & 02 27 38.33  & +35 28 43.8   & 164  & 147  & $-0.82$   \\
& B0222.3+3650  & 02 25 22.43  & +37 03 40.9   & 146  & 172  & $-0.95$   \\
& B0220.0+3704  & 02 23 07.86  & +37 18 03.6   & 200  & 198  & $-0.82$   \\
& B0224.8+3607A & 02 27 51.15  & +36 21 29.6   & 97   & 98   & $-1.23$   \\
& B0224.8+3607  & 02 27 52.76  & +36 21 01.3   & 97   & 168  & \nodata   \\
& B0224.8+3607B & 02 27 54.99  & +36 20 24.2   & 88   & 69   & $-0.79$   \\
& B0214.1+3430  & 02 17 05.91  & +34 44 30.4   & 128  & 136  & $-0.82$   \\
& B0211.5+3505  & 02 14 29.85  & +35 19 20.0   & 213  & 232  & \nodata   \\
& B0223.2+3441  & 02 26 17.12  & +34 54 36.5   & 88   & 85   & $-0.71$   \\
& B0224.7+3617  & 02 27 50.29  & +36 30 47.1   & 200  & 178  & $-0.82$   \\
& B0219.6+3415  & 02 22 38.27  & +34 29 29.1   & 188  & 174  & $-0.81$   \\
& B0211.2+3511  & 02 14 13.61  & +35 25 07.2   & 357  & 359  & $-0.91$   \\
& B0211.0+3518  & 02 14 01.38  & +35 32 28.7   & 174  & 189  & $-0.73$   \\
& B0211.9+3452  & 02 14 58.14  & +35 06 39.5   & 3011 & 3084 & $-0.83$   \\
& \nodata       & 02 14 57.959 & +35 06 32.27  & 259  & 869  & \nodata   \\
\tableline
$1.5\arcdeg-2\arcdeg$ & B0210.7+3533  & 02 13 43.25  & +35 47 29.5   & 149  & 141  & $-0.97$   \\
& B0214.3+3700  & 02 17 22.91  & +37 14 47.8   & 669  & 650  & $-1.18$   \\
& \nodata       & 02 17 22.726 & +37 14 47.08  & 207  & 248  & \nodata   \\
& B0222.3+3656  & 02 25 27.31  & +37 10 27.9   & 373  & 380  & $-0.45$   \\
& B0216.0+3413  & 02 19 01.20  & +34 27 45.7   & 131  & 136  & $-0.66$   \\
& B0225.0+3620  & 02 28 07.42  & +36 34 09.5   & 227  & 235  & $-0.83$   \\
& B0216.6+3409  & 02 19 37.82  & +34 23 10.5   & 555  & 515  & $-1.06$   \\
& B0215.1+3710  & 02 18 11.43  & +37 24 36.5   & 481  & 473  & $-0.70$   \\
& \nodata       & 02 18 11.421 & +37 24 35.44  & 184  & 192  & \nodata   \\
& B0223.9+3646  & 02 26 59.91  & +36 59 27.9   & 76   & 60   & $-0.75$   \\
& B0219.2+3717  & 02 22 15.42  & +37 31 16.7   & 144  & 123  & $-0.08$   \\
& B0225.9+3602  & 02 28 57.74  & +36 16 13.8   & 125  & 122  & $-0.80$   \\
& B0212.8+3429  & 02 15 49.40  & +34 43 02.8   & 213  & 199  & $-0.19$   \\
& B0221.9+3417  & 02 24 56.18  & +34 30 35.8   & 1657 & 1638 & $-0.91$   \\
& \nodata       & 02 24 56.629 & +34 30 27.43  & 184  & 222  & \nodata   \\
& B0211.1+3634  & 02 14 11.25  & +36 48 24.9   & 228  & 244  & $-1.20$   \\
& B0216.2+3405B & 02 19 18.40  & +34 19 49.5   & 273  & 338  & $-0.96$   \\
& B0216.2+3405  & 02 19 15.49  & +34 19 42.1   & 412  & 802  & \nodata   \\
& B0216.2+3405A & 02 19 13.44  & +34 19 37.2   & 412  & 463  & $-0.86$   \\
& B0218.8+3402  & 02 21 50.01  & +34 16 19.1   & 359  & 347  & $-0.88$   \\
& B0226.2+3534  & 02 29 20.00  & +35 47 40.2   & 167  & 163  & $-0.80$   \\
& B0224.7+3442  & 02 27 43.89  & +34 56 23.1   & 100  & 85   & $-0.90$   \\
& B0221.6+3406A & 02 24 34.48  & +34 21 39.5   & 104  & 109  & $-1.11$   \\
& B0209.5+3536  & 02 12 31.09  & +35 50 40.1   & 381  & 482  & \nodata   \\
& B0226.1+3509  & 02 29 11.30  & +35 22 22.6   & 160  & 178  & \nodata   \\
& B0224.2+3430  & 02 27 16.40  & +34 43 42.9   & 129  & 128  & $-0.96$   \\
& B0221.3+3405  & 02 24 20.16  & +34 19 17.5   & 195  & 193  & $-0.84$   \\
& B0221.6+3406  & 02 24 39.96  & +34 20 03.1   & 459  & 543  & \nodata   \\
& B0221.6+3406B & 02 24 41.08  & +34 19 43.6   & 459  & 433  & $-0.93$   \\
& B0220.7+3401  & 02 23 47.53  & +34 15 08.3   & 403  & 421  & $-0.80$   \\
& B0226.3+3619  & 02 29 24.87  & +36 32 26.6   & 191  & 174  & $-0.85$   \\
& B0224.9+3650  & 02 28 01.68  & +37 03 36.3   & 1409 & 1455 & $-0.90$   \\
& B0226.5+3618  & 02 29 35.60  & +36 31 22.9   & 417  & 401  & $-0.67$   \\
& \nodata       & 02 29 35.650 & +36 31 24.65  & 294  & 335  & \nodata   \\
& B0218.9+3353  & 02 21 56.58  & +34 07 17.5   & 127  & 122  & $-0.92$   \\
& B0216.9+3732  & 02 19 59.02  & +37 45 56.1   & 373  & 362  & $-0.78$   \\
& B0223.1+3408  & 02 26 10.15  & +34 21 30.7   & 3793 & 3678 & $-0.16$   \\
& \nodata       & 02 26 10.338 & +34 21 30.28  & 700  & 1834 & \nodata   \\
& B0212.4+3714  & 02 15 31.32  & +37 28 43.3   & 333  & 323  & $-0.71$   \\
& B0209.2+3504  & 02 12 14.34  & +35 18 24.7   & 253  & 230  & $-0.85$   \\
& B0225.1+3659  & 02 28 13.67  & +37 12 57.1   & 1566 & 1519 & $-1.09$   \\
& \nodata       & 02 28 13.706 & +37 12 58.23  & 643  & 692  & \nodata   \\
& B0208.4+3547  & 02 11 28.80  & +36 01 26.2   & 446  & 431  & $-0.94$   \\
& B0210.3+3652  & 02 13 22.53  & +37 07 00.0   & 303  & 320  & \nodata   \\
& B0225.6+3655  & 02 28 42.20  & +37 08 58.0   & 306  & 308  & $-0.85$   \\
& B0225.6+3655A & 02 28 42.07  & +37 09 03.6   & 306  & 286  & $-0.80$   \\
& B0212.5+3405  & 02 15 29.49  & +34 19 47.5   & 337  & 328  & $-0.86$   \\
\tableline
$2\arcdeg-3\arcdeg$ & B0218.6+3342  & 02 21 40.00  & +33 56 36.1   & 195  & 204  & $-0.84$   \\
& B0219.8+3343A & 02 22 48.79  & +33 57 18.9   & 638  & 638  & $-0.83$   \\
& B0219.8+3343  & 02 22 48.85  & +33 57 11.3   & 638  & 688  & $-0.88$   \\
& B0228.1+3547  & 02 31 11.57  & +36 00 28.4   & 321  & 303  & $-1.33$   \\
& B0219.2+3339  & 02 22 16.75  & +33 53 27.3   & 516  & 499  & $-0.67$   \\
& \nodata       & 02 22 16.801 & +33 53 26.98  & 331  & 337  & \nodata   \\
& B0225.5+3419  & 02 28 34.05  & +34 33 09.5   & 504  & 568  & $-1.06$   \\
& B0225.5+3419B & 02 28 34.63  & +34 32 58.5   & 504  & 492  & $-0.97$   \\
& B0216.8+3334  & 02 19 48.77  & +33 48 16.0   & 1352 & 1357 & $-0.81$   \\
& B0212.6+3736  & 02 15 40.01  & +37 50 09.3   & 691  & 655  & $-0.54$   \\
& B0223.9+3351  & 02 26 55.50  & +34 04 55.7   & 532  & 525  & $-1.02$   \\
& B0208.9+3429  & 02 11 54.12  & +34 44 01.7   & 569  & 551  & $-1.08$   \\
& B0210.5+3404  & 02 13 28.47  & +34 18 20.0   & 402  & 482  & $-0.90$   \\
& B0216.0+3756B & 02 19 09.03  & +38 10 00.7   & 439  & 645  & $-0.84$   \\
& B0216.0+3756  & 02 19 08.14  & +38 10 15.6   & 439  & 739  & \nodata   \\
& B0206.6+3533  & 02 09 38.91  & +35 47 48.9   & 4259 & 5519 & $-0.67$   \\
& B0206.6+3533A & 02 09 38.87  & +35 47 48.9   & 4259 & 5489 & $-0.66$   \\
& B0213.2+3750  & 02 16 19.76  & +38 04 39.4   & 709  & 782  & $-0.89$   \\
& B0229.4+3517  & 02 32 29.32  & +35 30 53.6   & 2149 & 2359 & $-0.86$   \\
& B0227.3+3713  & 02 30 25.65  & +37 26 18.0   & 1034 & 1013 & $-1.10$   \\
& B0214.7+3321  & 02 17 43.31  & +33 35 03.0   & 683  & 657  & $-0.85$   \\
& B0228.9+3429  & 02 31 59.53  & +34 42 26.7   & 2361 & 2556 & $-1.02$   \\
& B0228.1+3729  & 02 31 14.98  & +37 42 57.0   & 1426 & 1397 & $-1.03$   \\
& B0229.4+3410  & 02 32 28.67  & +34 24 03.0   & 9334 & 9459 & $-0.88$   
\enddata
\tablecomments{Units of right ascension are hours, minutes and seconds, and units of declination are degree, arcminutes and arcseconds. The first data entry for each target refers to the WENSS target source, and the second entry, when present, refers to the VLBI detection. The spectral index $\alpha$ is defined as $S_{\nu}\propto\nu^{\alpha}$ and is estimated from WENSS and NVSS integrated flux densities where available.}
\label{tab:tabf2}
\end{deluxetable}

\clearpage
\begin{figure}
\epsscale{1.0}
\plotone{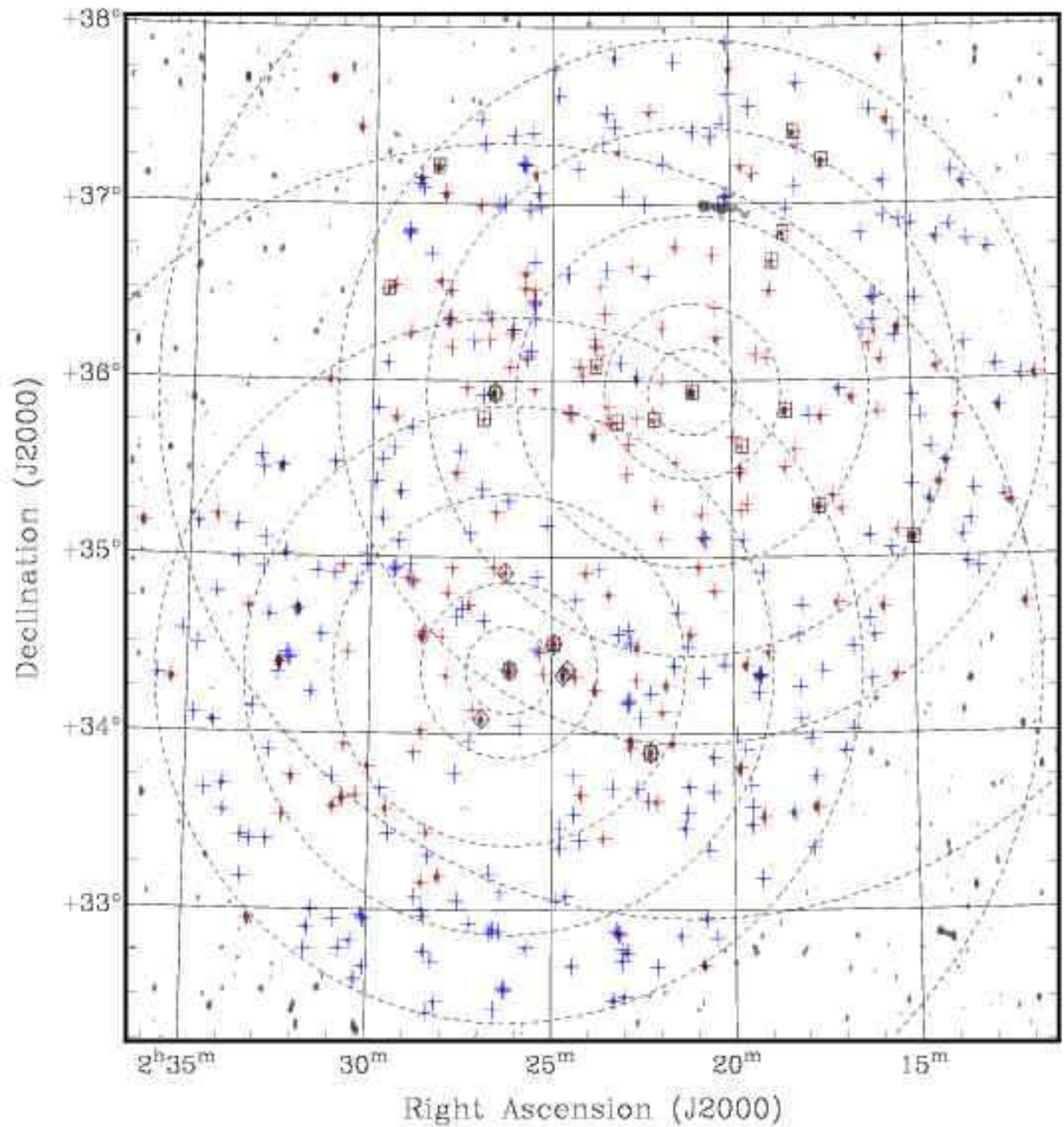}
\caption{Gray-scale WENSS image of the two fields surveyed by the VLBA, JB and WB VLBI observation. The dashed circles define the six sub-fields that are co-located on the phase centre of each field of the VLBI observation. The crosses located across the image denote all WENSS sources that were targeted by the VLBI observation, red crosses mark unresolved WENSS sources with peak flux densities above the VLBI sensitivity limit and green crosses mark sources that were resolved in WENSS or fell below the VLBI sensitivity limit. Targets that are boxed identify VLBI sources detections in the B0218$+$357 field and those that are contained within a diamond identify VLBI detections in the J0226$+$3421 field. The four sources that were commonly detected within both of the fields and are marked within a box and diamond.}
\label{fig:figfield}            
\end{figure}

\clearpage
\begin{figure}
\epsscale{0.4}
\begin{center}
\mbox{
\plotone{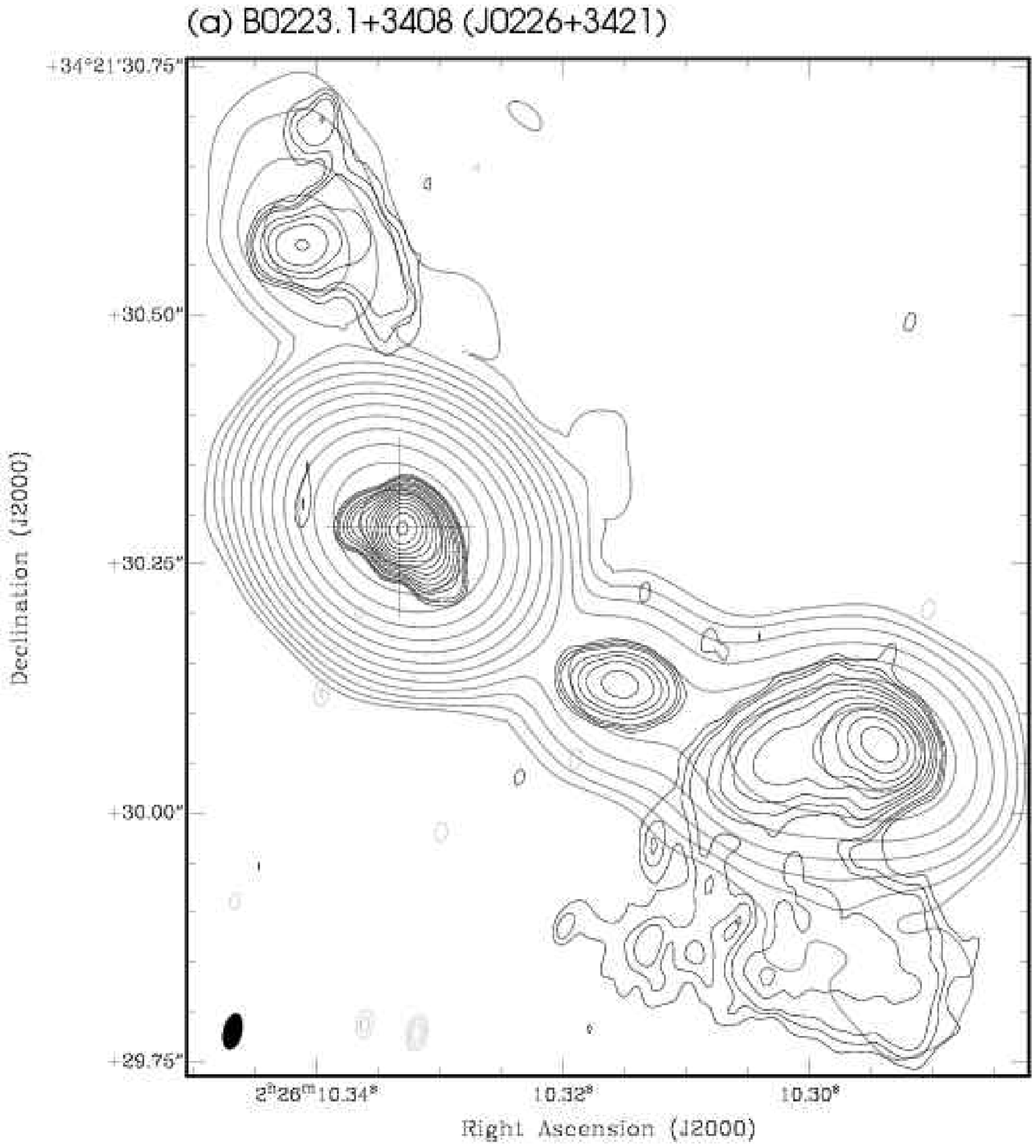} \quad
\plotone{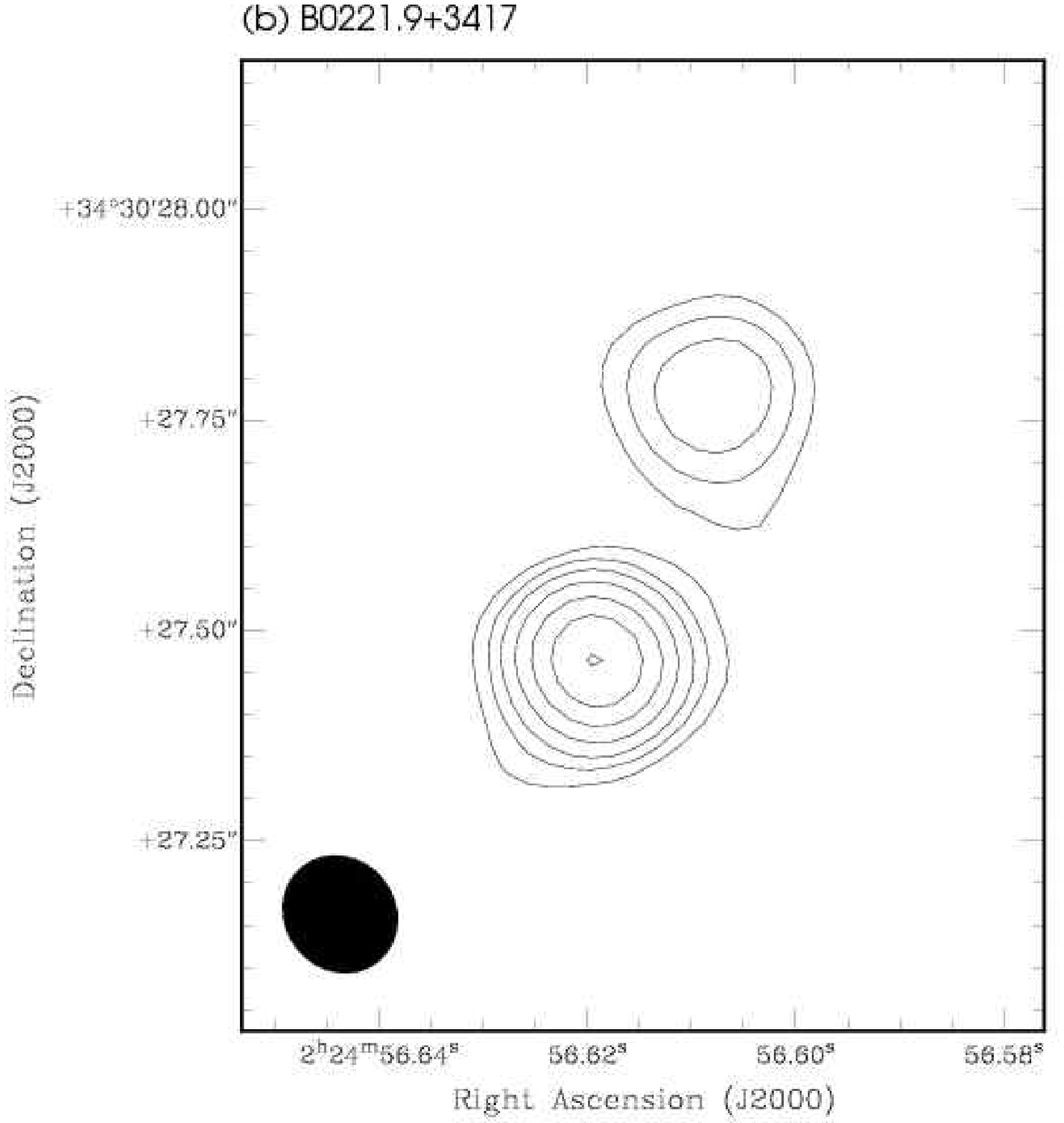}
}
\mbox{
\plotone{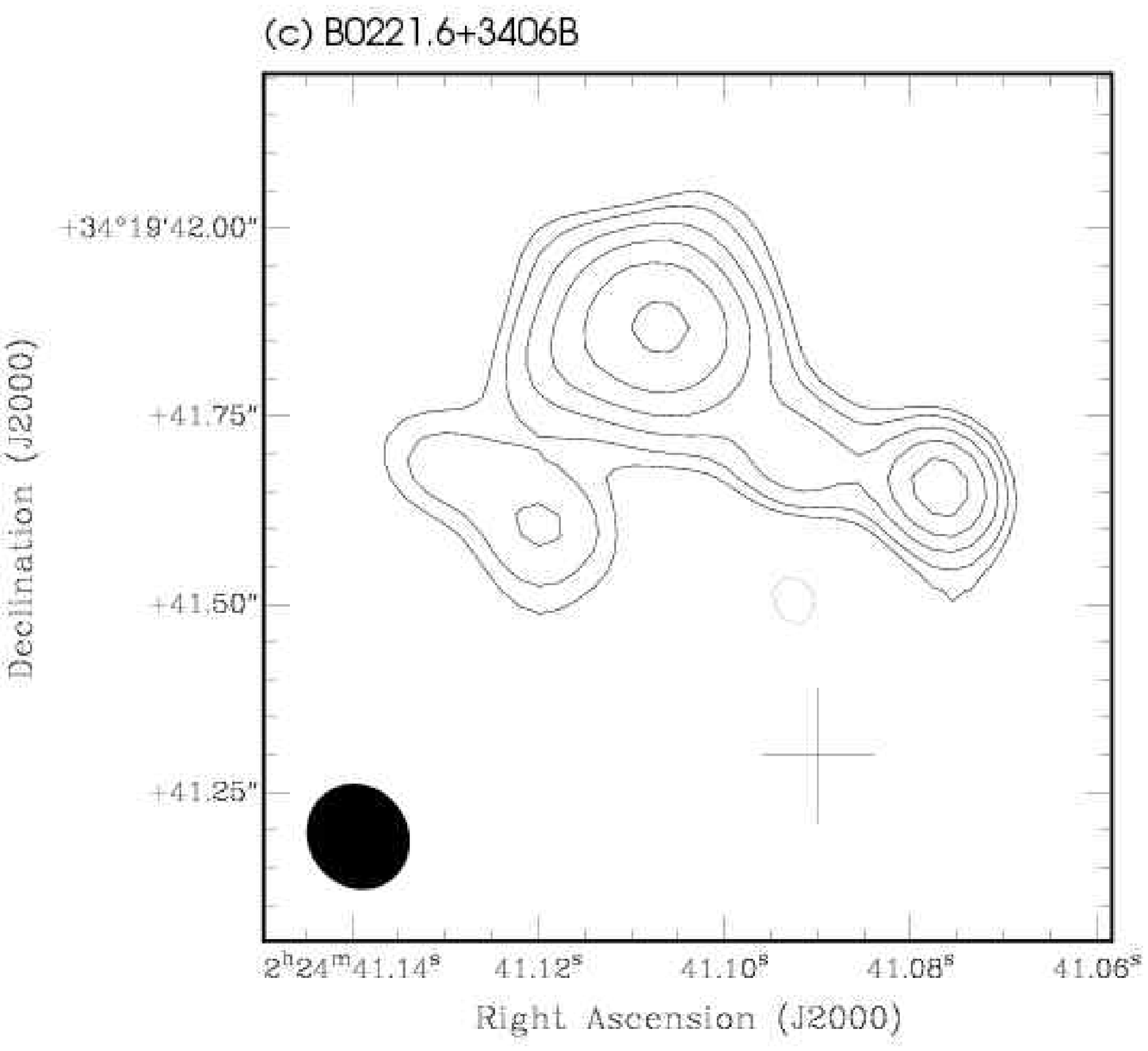} \quad
\plotone{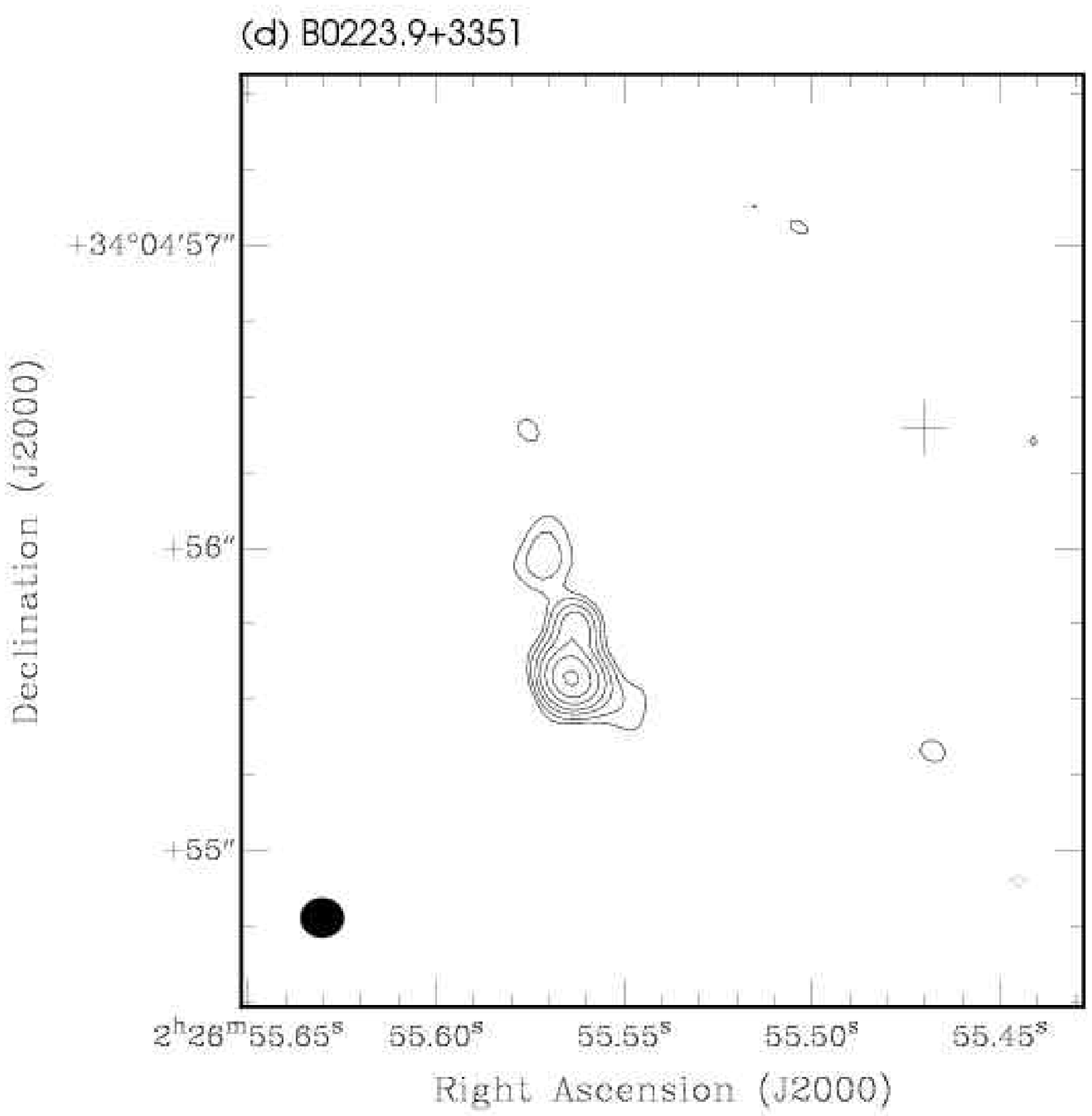}
}
\caption{Naturally weighted VLBI images of field 1 sources. Contours are drawn at $\pm2^{0}, \pm2^{\frac{1}{2}}, \pm2^{1}, \pm2^{\frac{3}{2}}, \cdots$ times the $3\sigma$ rms noise except for B0223.1$+$3408 where the lowest contour is at $1.5\sigma$ rms noise. Restoring beam and rms image noise for all images can be found in Table \ref{tab:tabastrometry}. Crosses mark the best known radio positions (see Table \ref{tab:tabastrometry}). Notes for sub-figures (a) Grey contours: 2 cm VLA $+$ Pie Town contour map; $1\sigma$ RMS noise is 0.053 mJy beam$^{-1}$; contours at $\pm1, \pm2, \pm4, \pm8, \cdots$ times the $3\sigma$ rms noise; beam size is $118\times96$ mas at P.A.$=51\arcdeg$; (h) Grey contours: Field 2 detection of source restored using the same beam as the field 1 source.}
\label{fig:figf1}
\end{center}
\end{figure}
\clearpage
\epsscale{0.4}
\begin{center}
\mbox{
\plotone{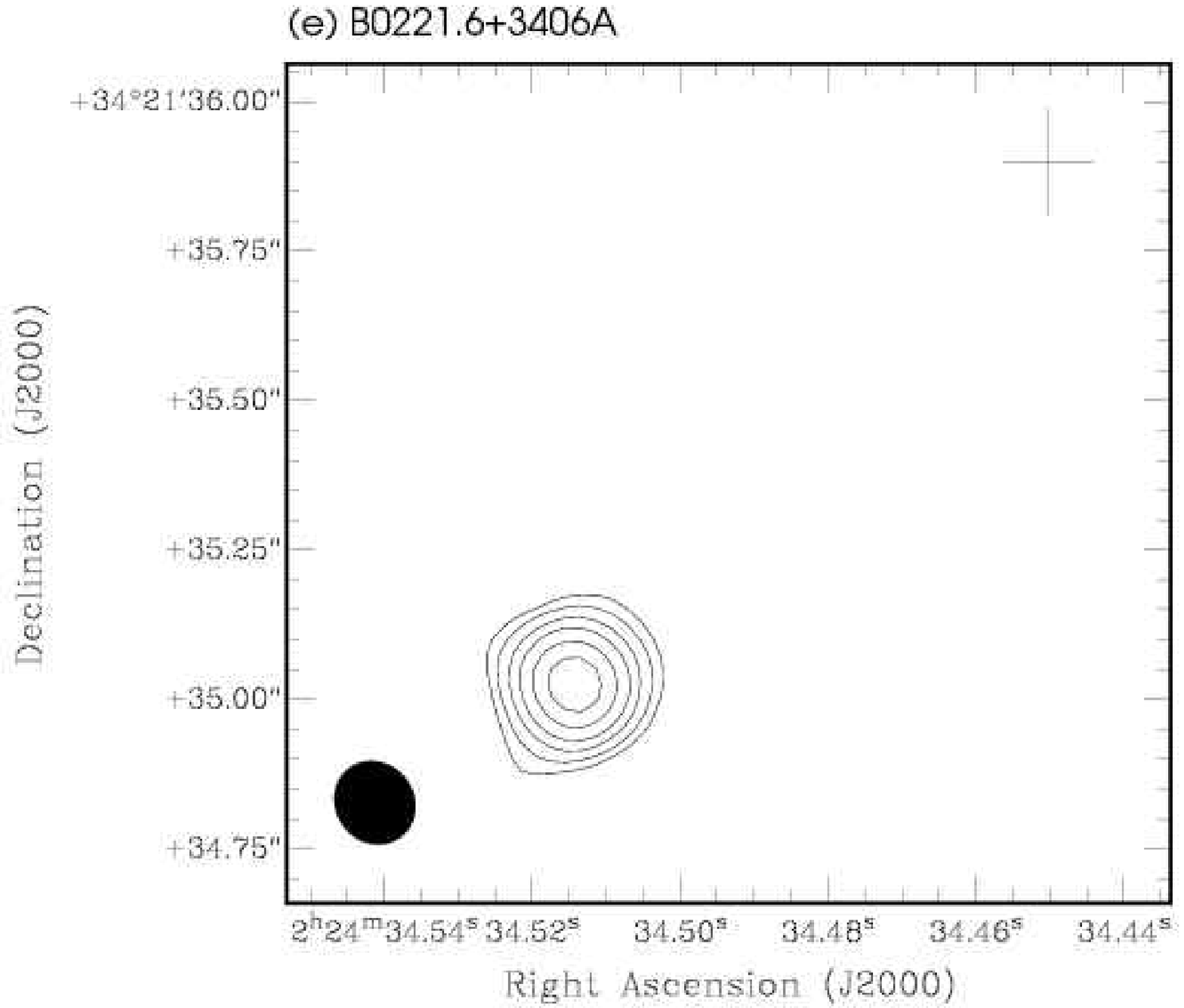} \quad
\plotone{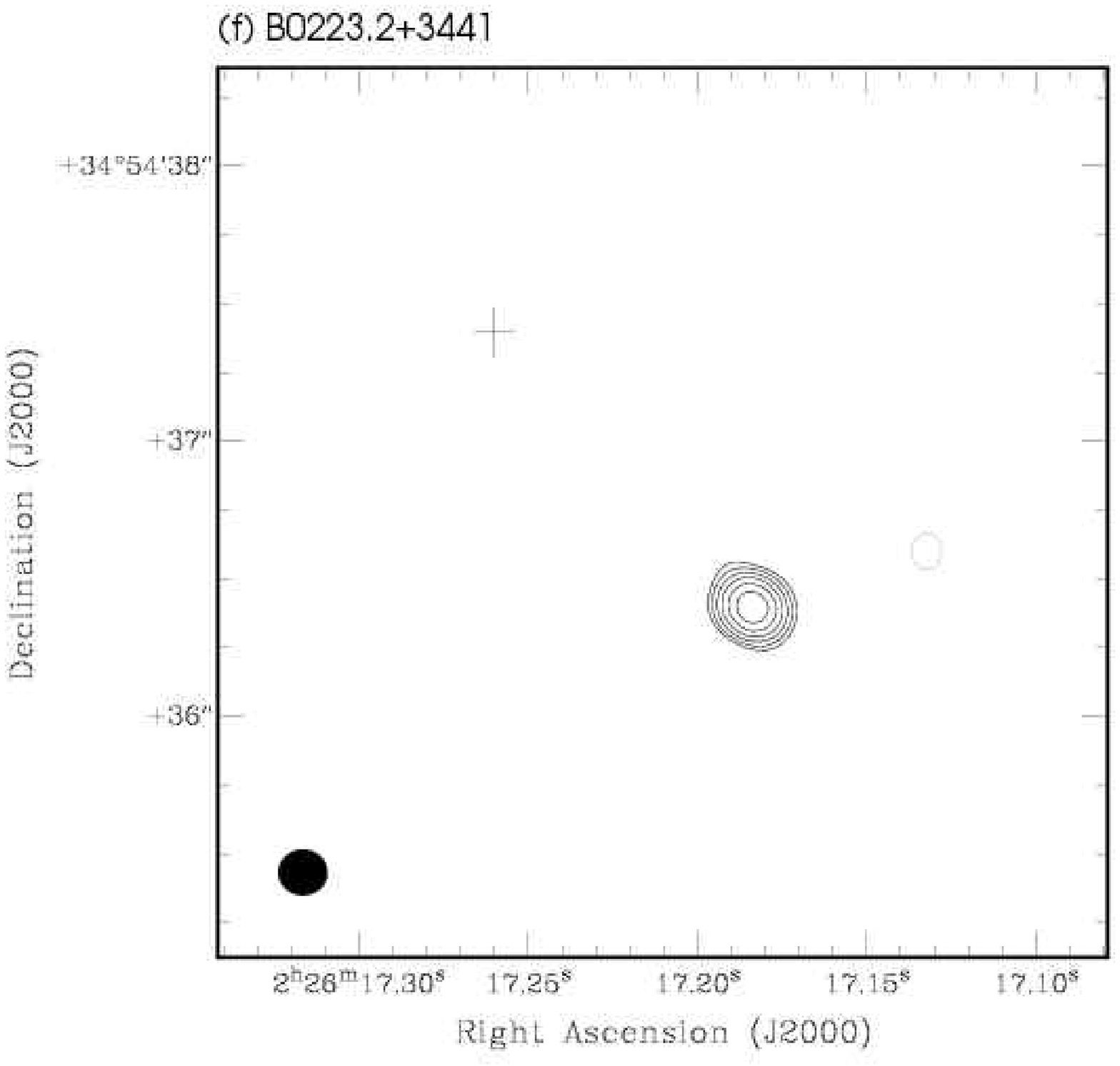}
}\\[5mm]
\mbox{
\plotone{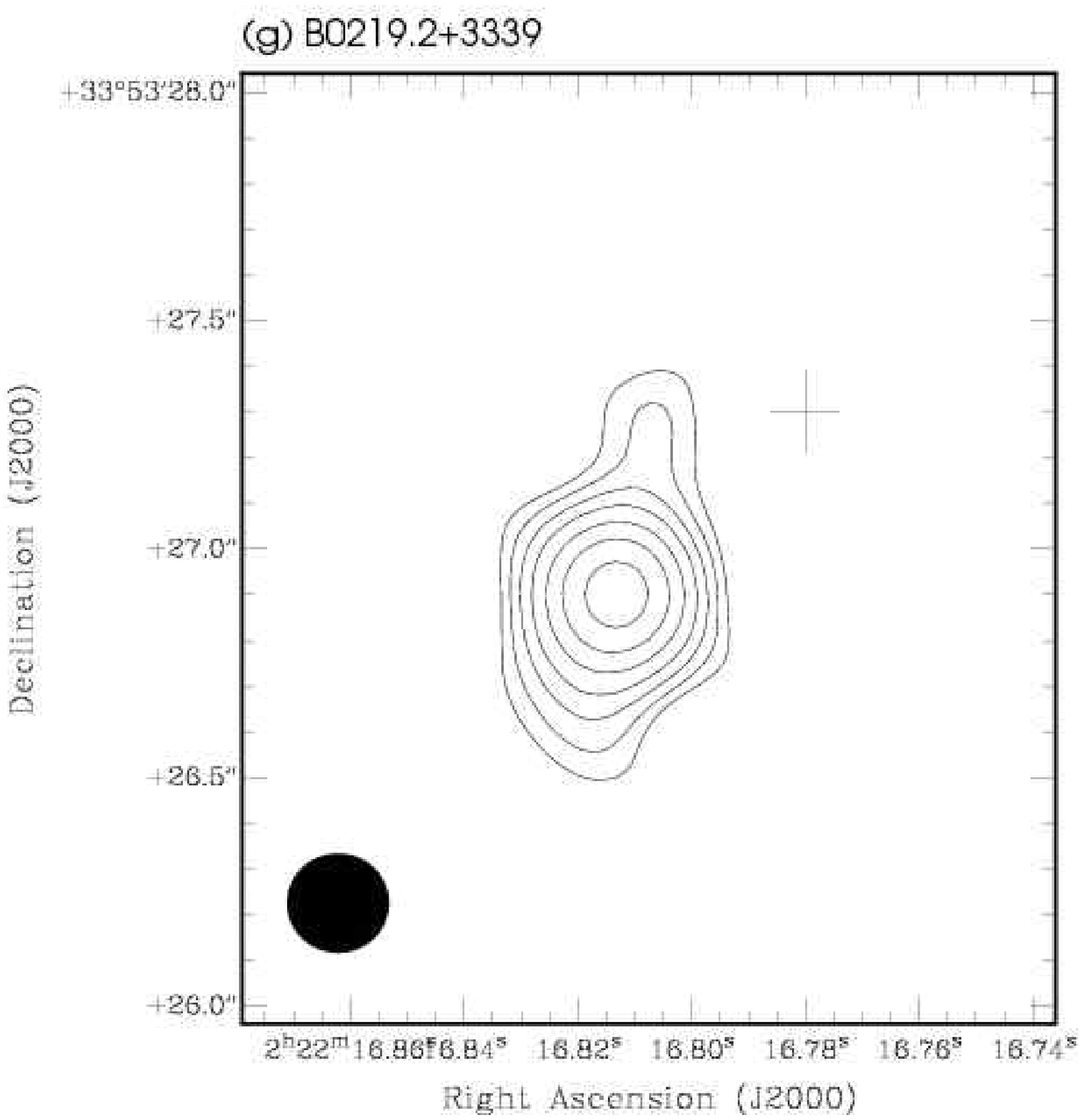} \quad
\plotone{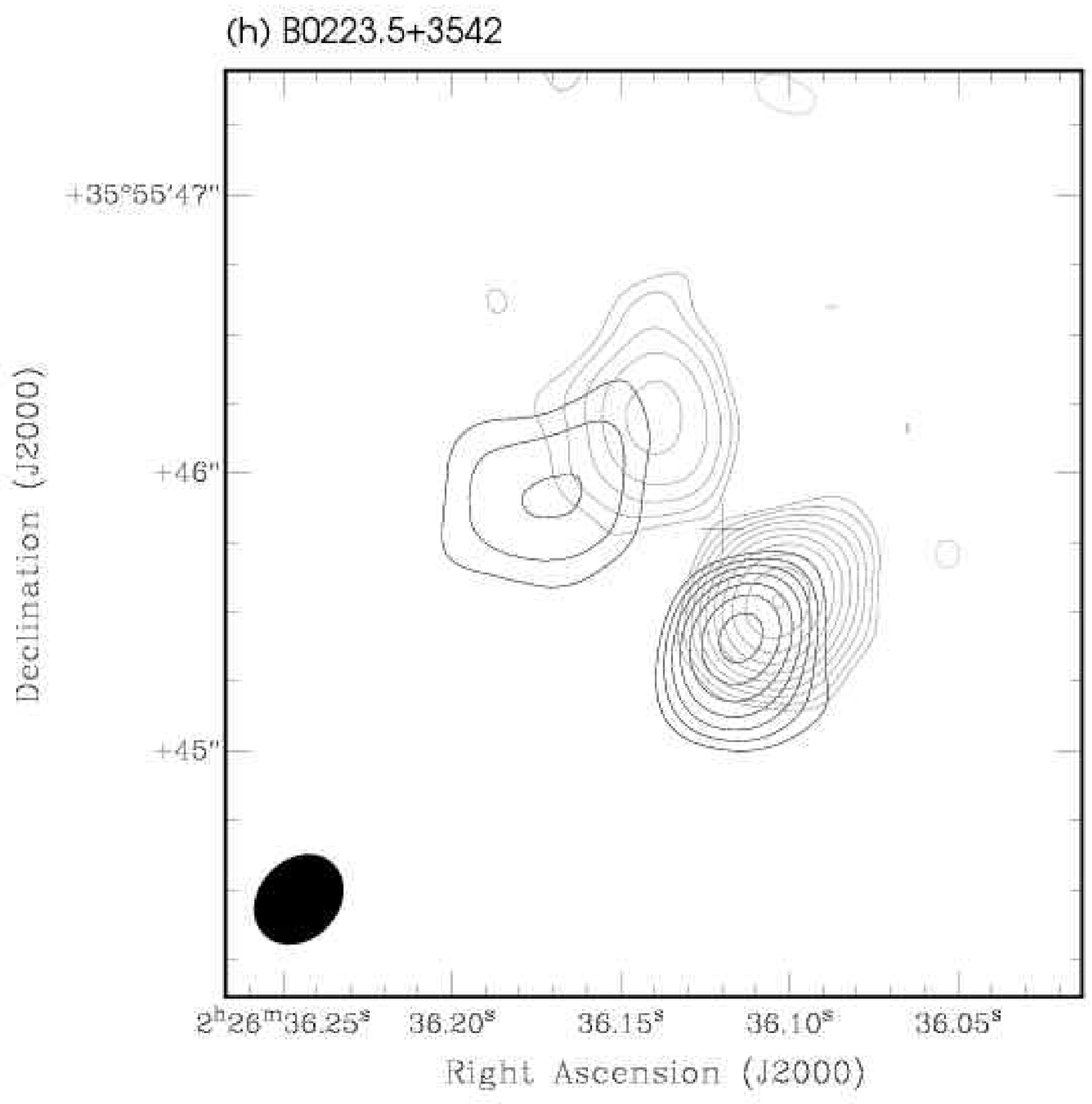}
}\\[5mm]
{Fig. 2. --- Continued}
\end{center}
\clearpage

\begin{figure}
\epsscale{0.4}
\begin{center}
\mbox{
\plotone{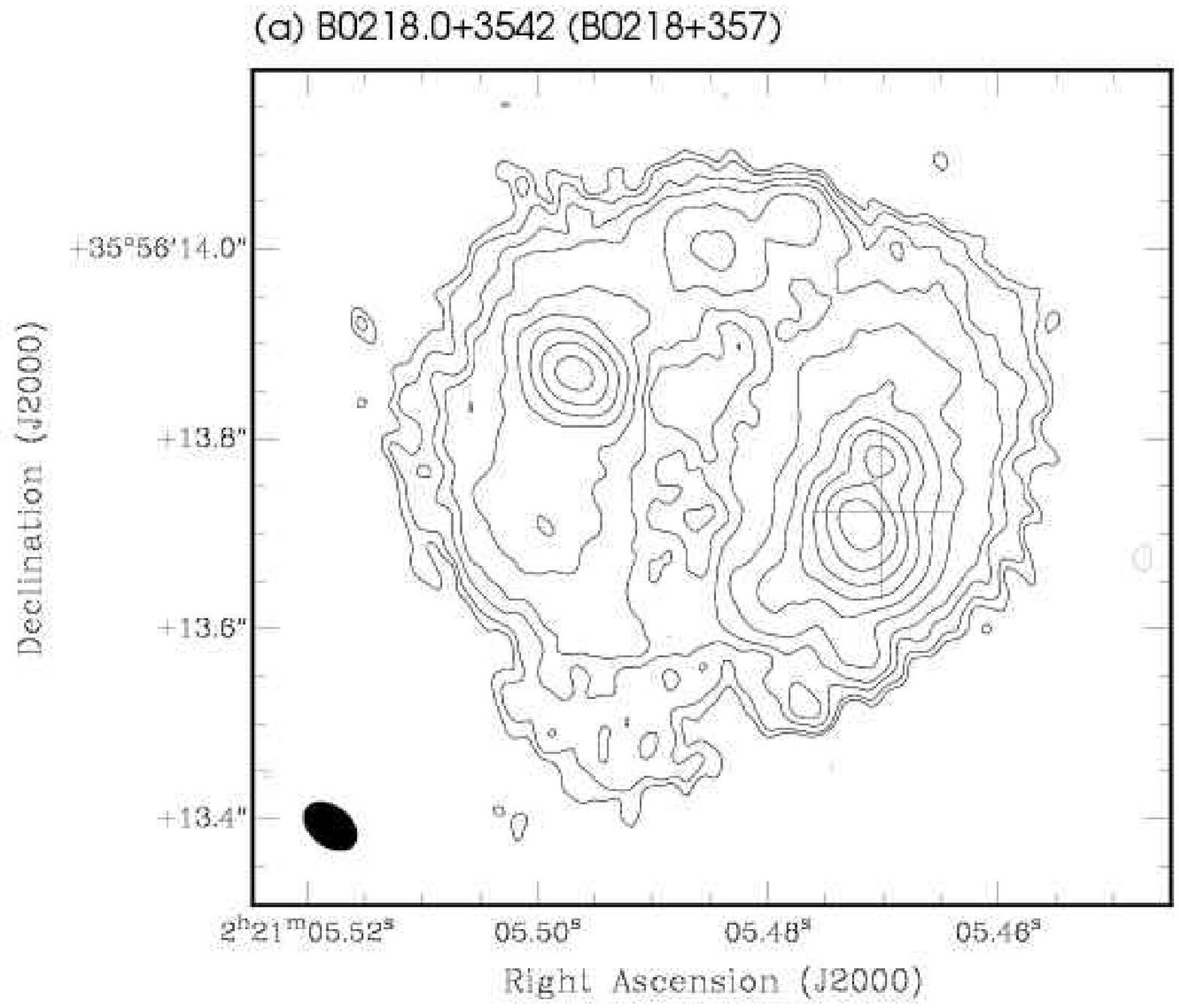} \quad
\plotone{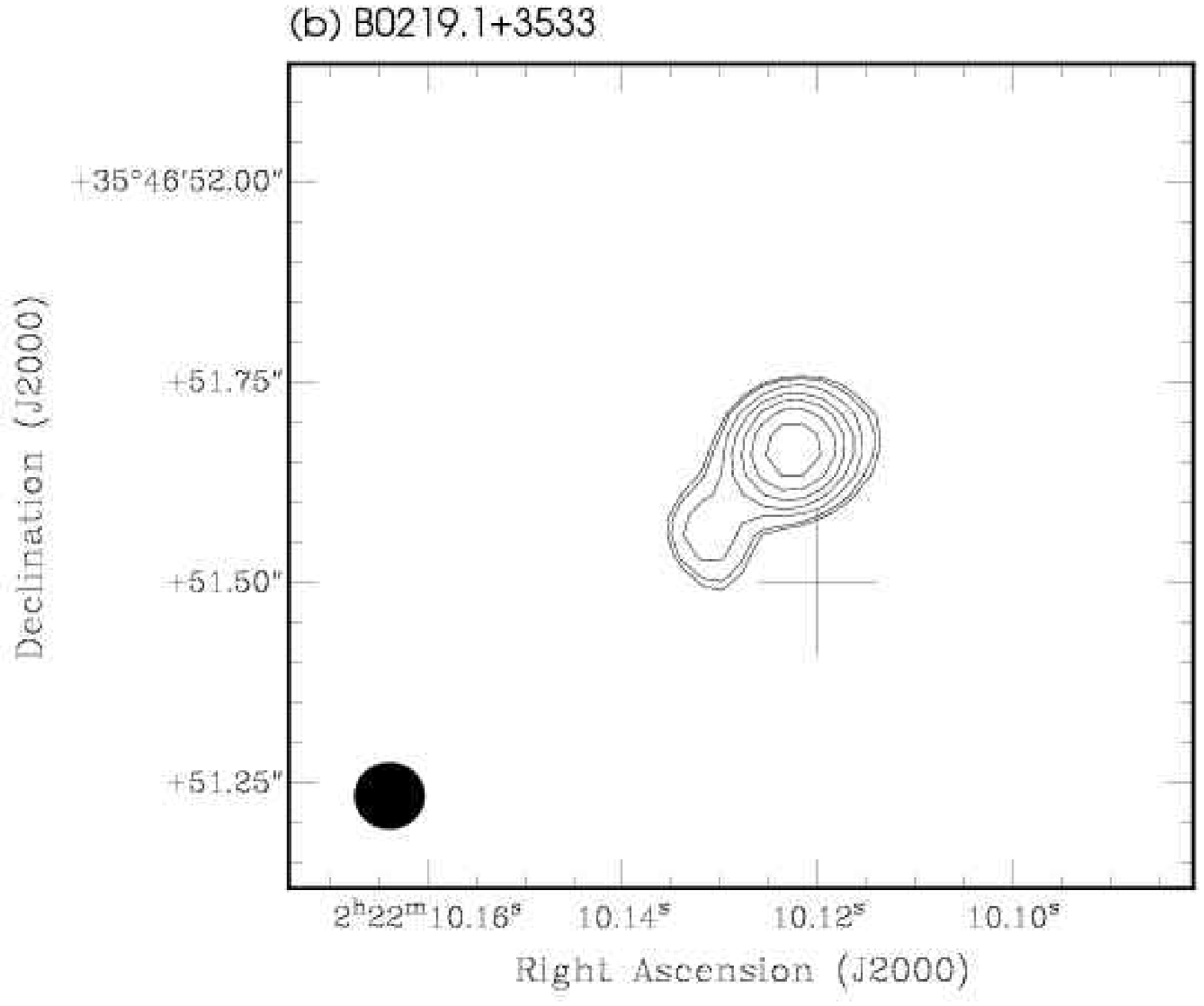}
}
\mbox{
\plotone{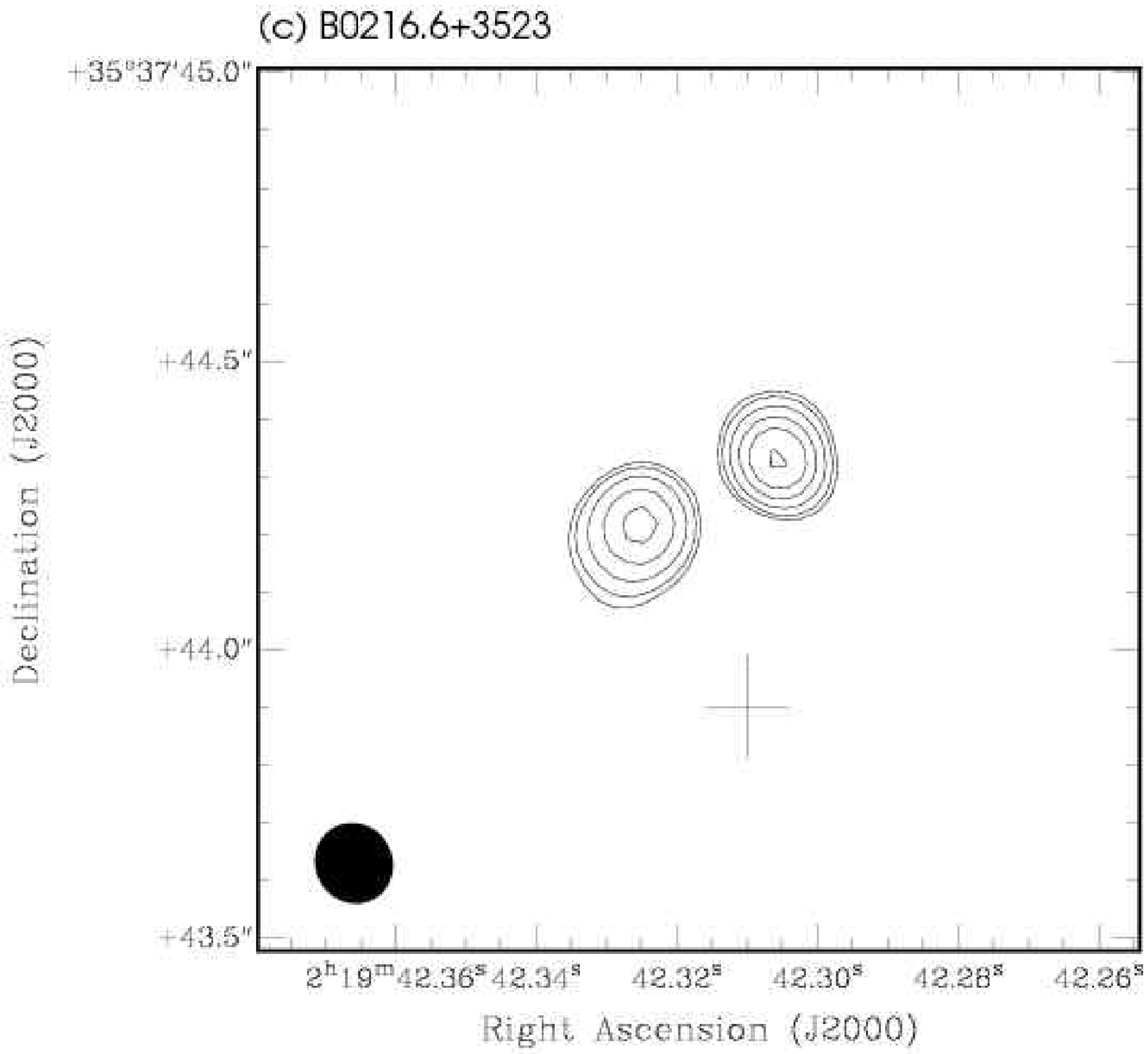} \quad
\plotone{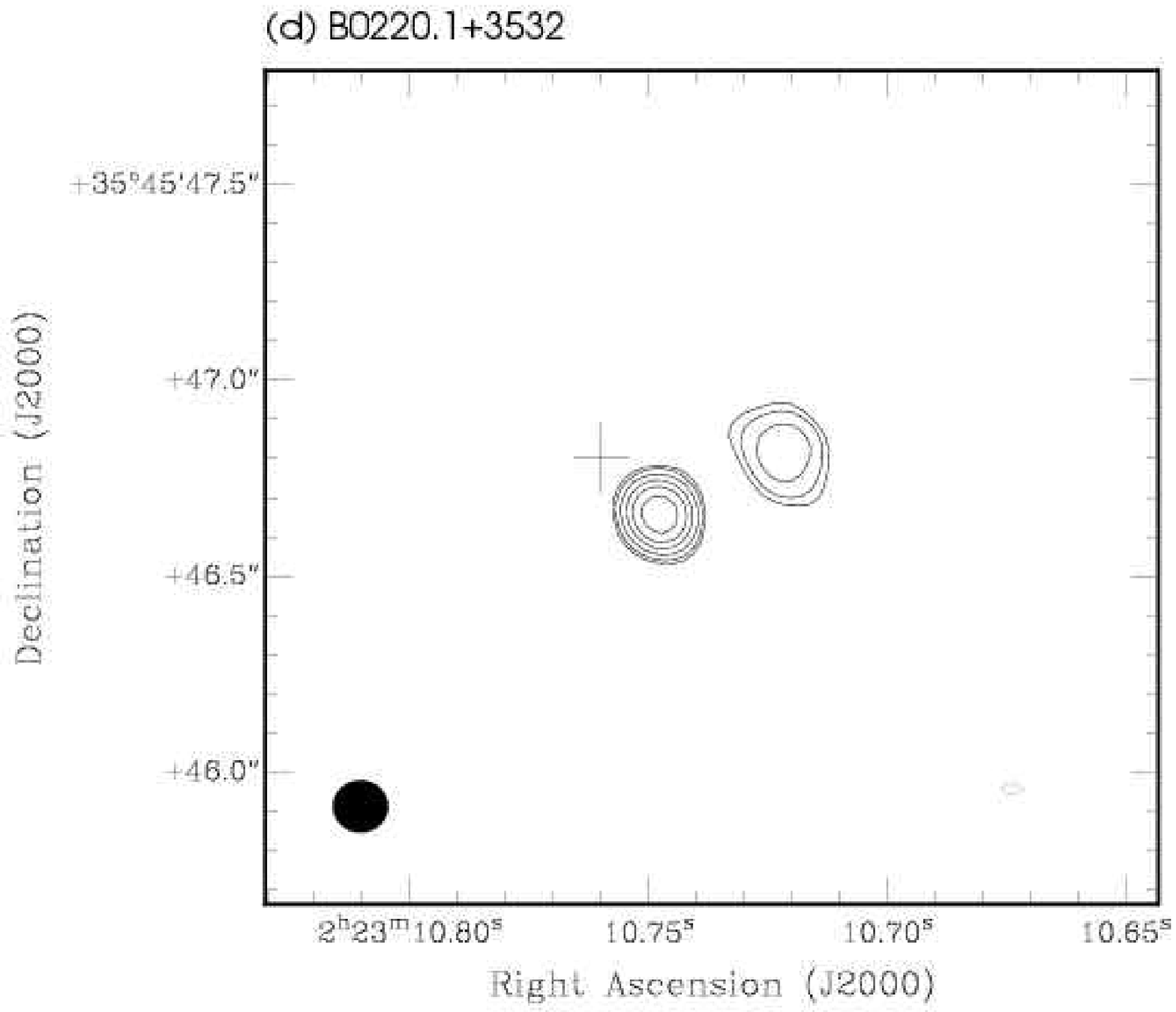}
}
\caption{Naturally weighted VLBI images of field 2 sources. Contours are drawn at $\pm2^{0}, \pm2^{\frac{1}{2}}, \pm2^{1}, \pm2^{\frac{3}{2}}, \cdots$ times the $3\sigma$ rms noise. Restoring beam and rms image noise for all images can be found in Table \ref{tab:tabastrometry}. Crosses mark the best known radio positions (see Table \ref{tab:tabastrometry}). Notes for sub-figures (e) and (h) Grey contours: 1465 MHz VLA contour map \citep{rot94}; (o), (q) and (s) Grey contours: Field 1 detection of source restored using the same beam as the field 2 source.}
\label{fig:figf2}            
\end{center}
\end{figure}
\clearpage
\epsscale{0.4}
\begin{center}
\mbox{
\plotone{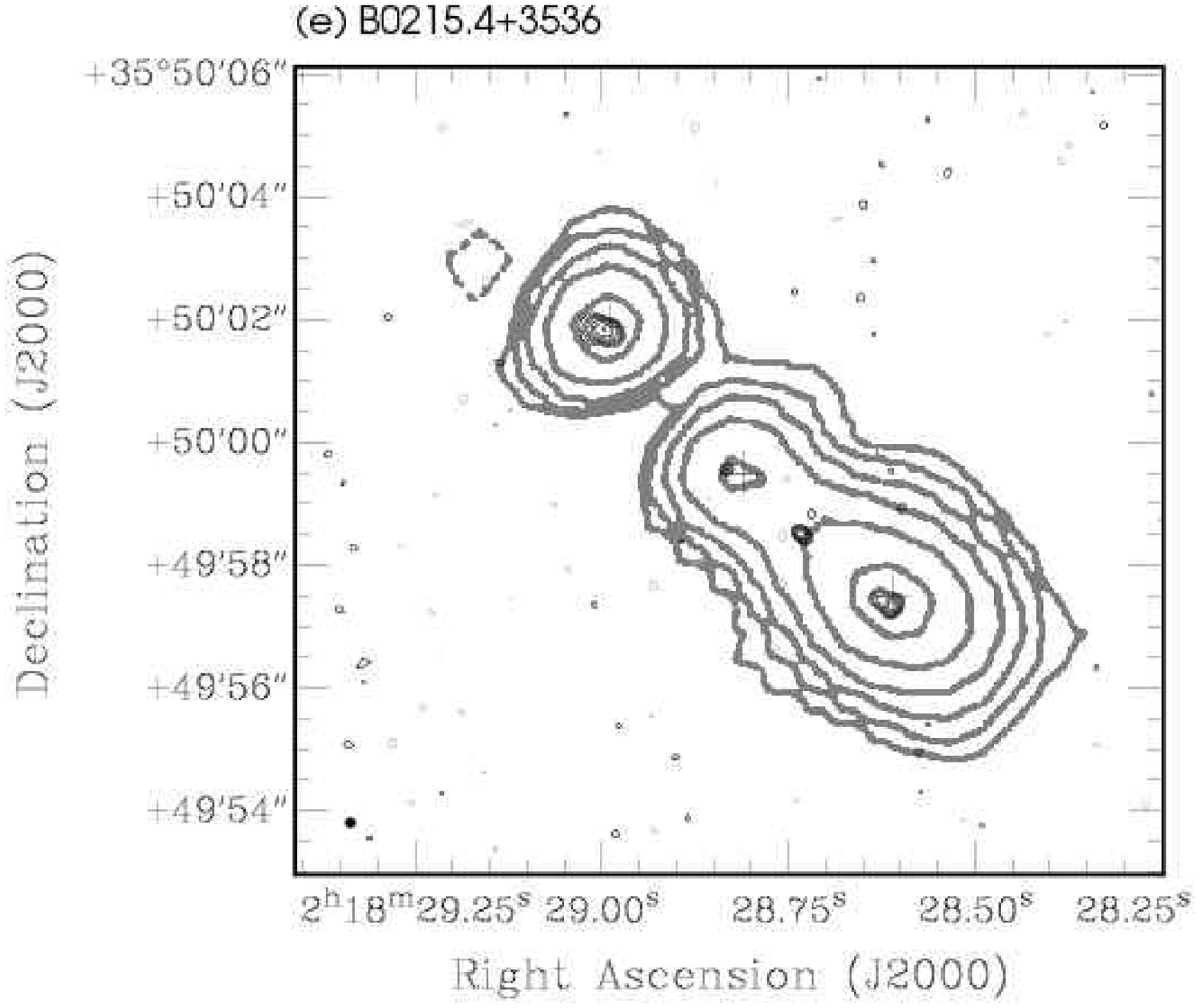} \quad
\plotone{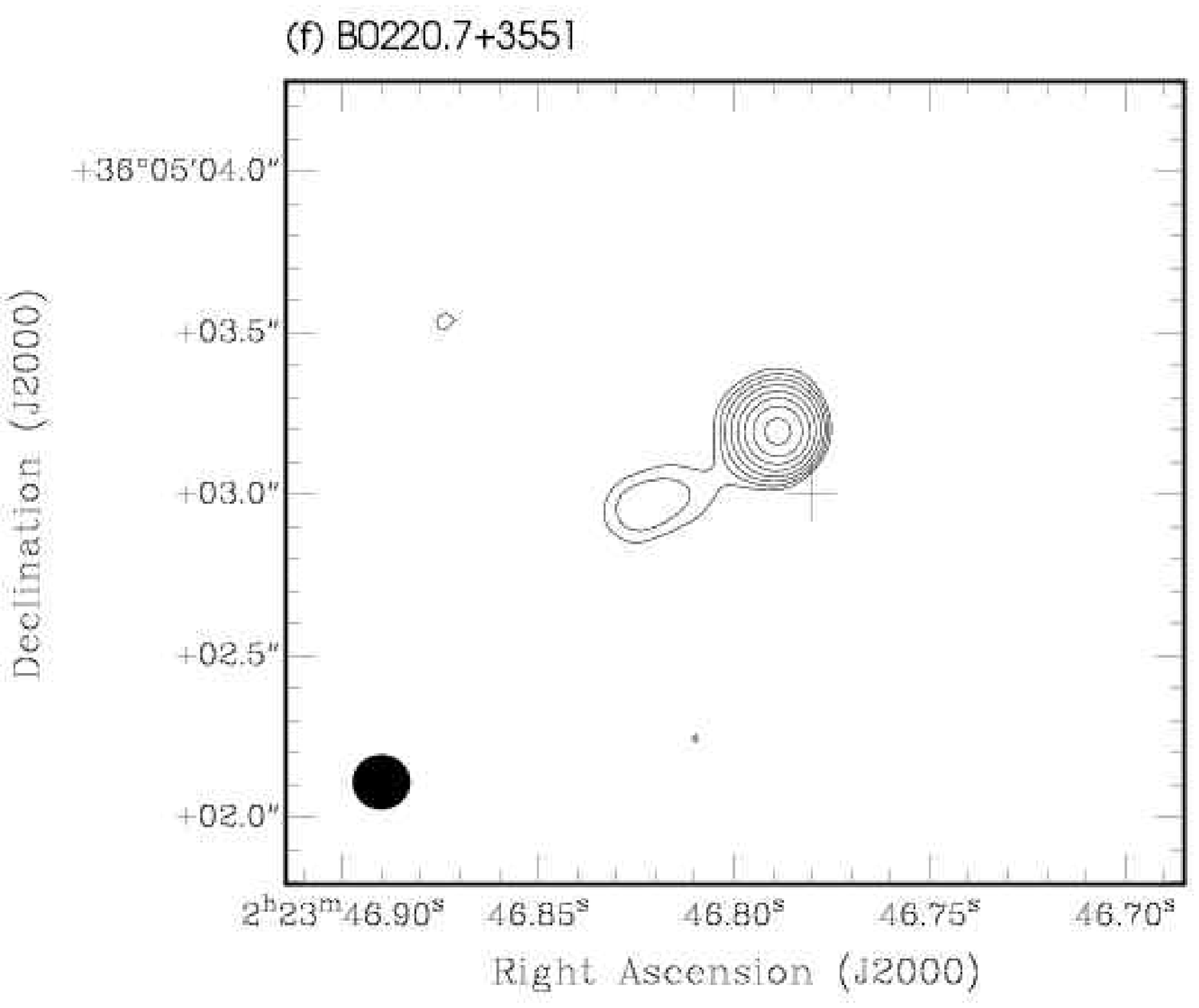}
}\\[5mm]
\mbox{
\plotone{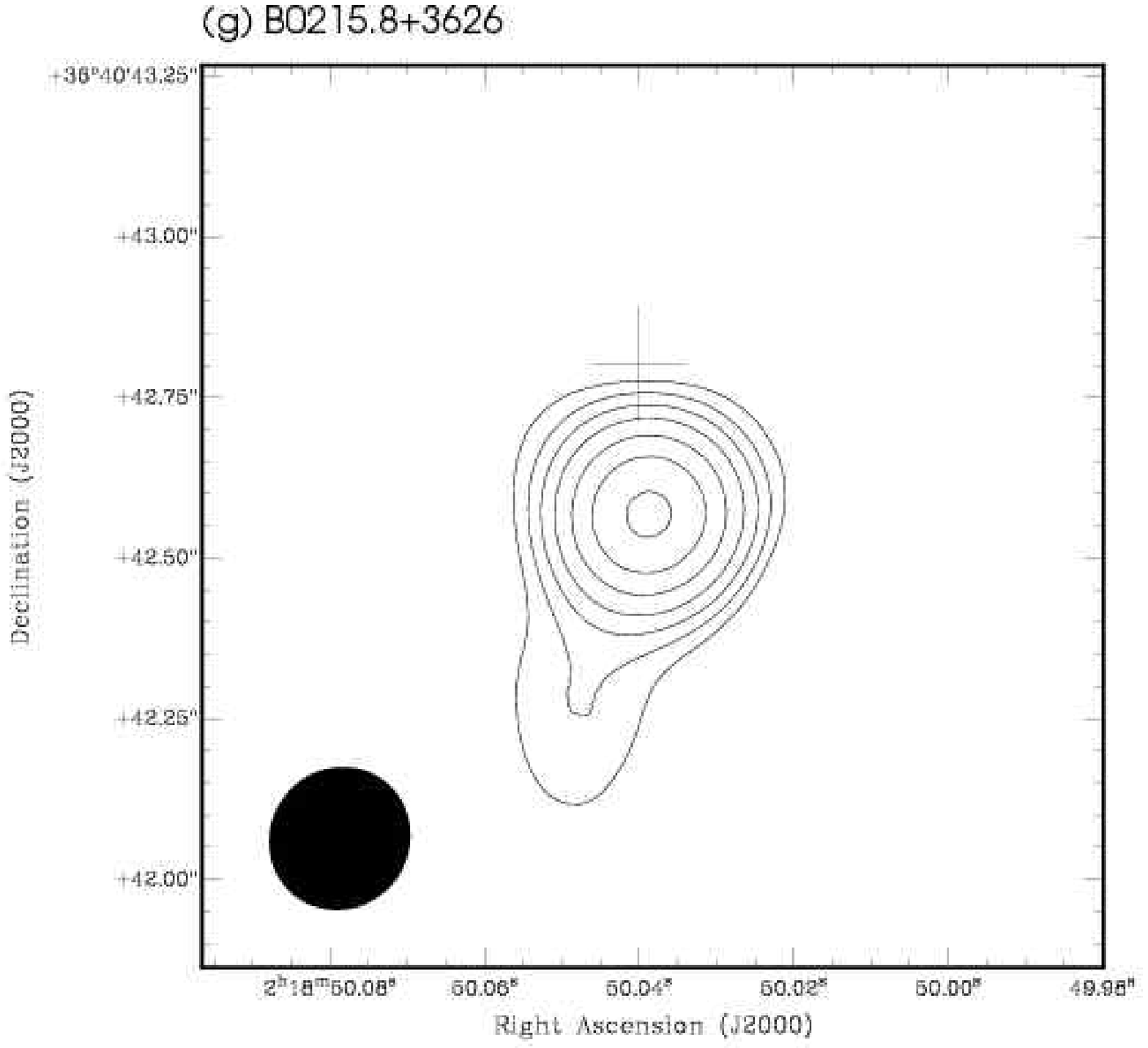} \quad
\plotone{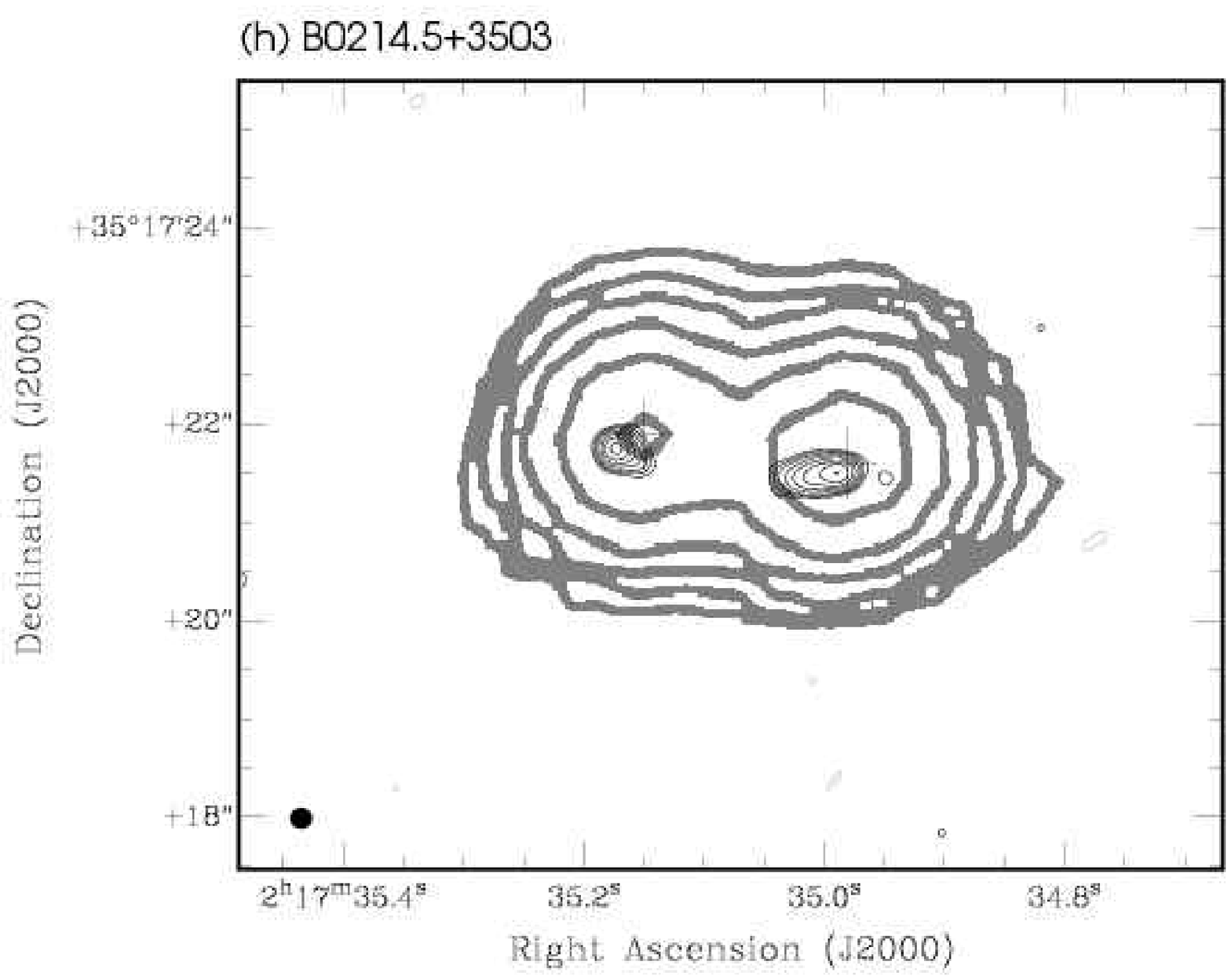}
}\\[5mm]
\mbox{
\plotone{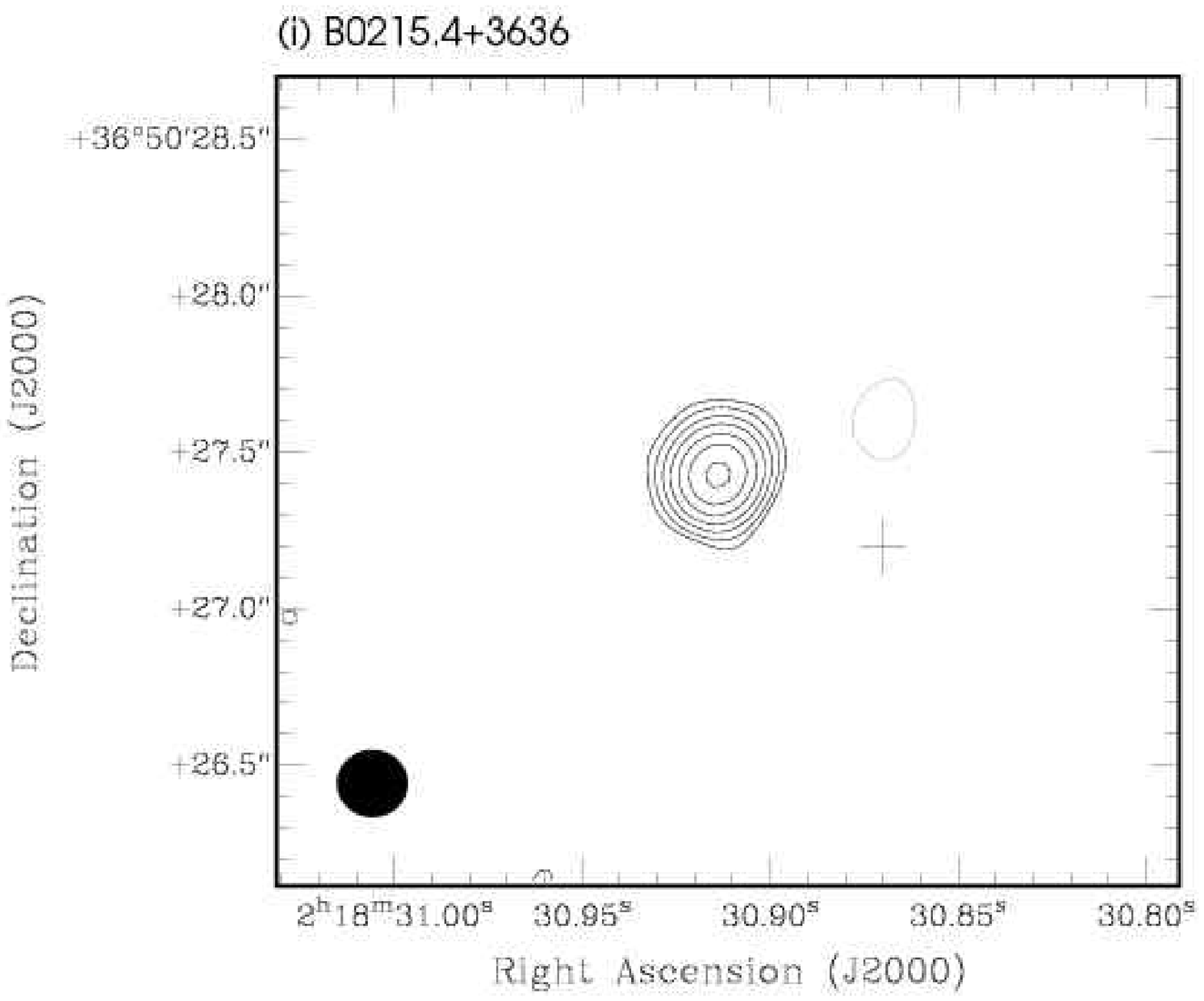} \quad
\plotone{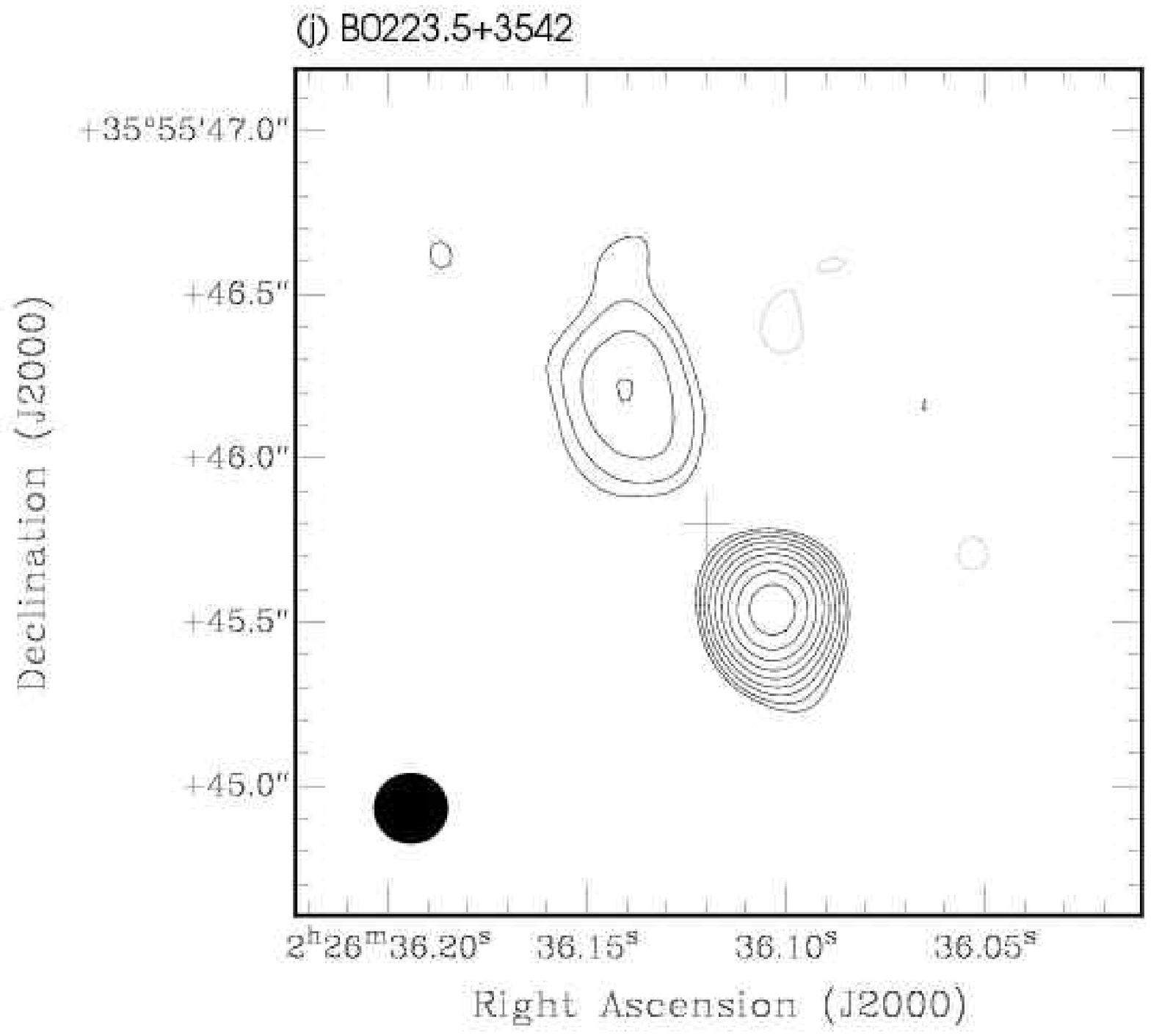}
}\\[5mm]
{Fig. 3. --- Continued}
\end{center}
\clearpage
\epsscale{0.38}
\begin{center}
\mbox{
\plotone{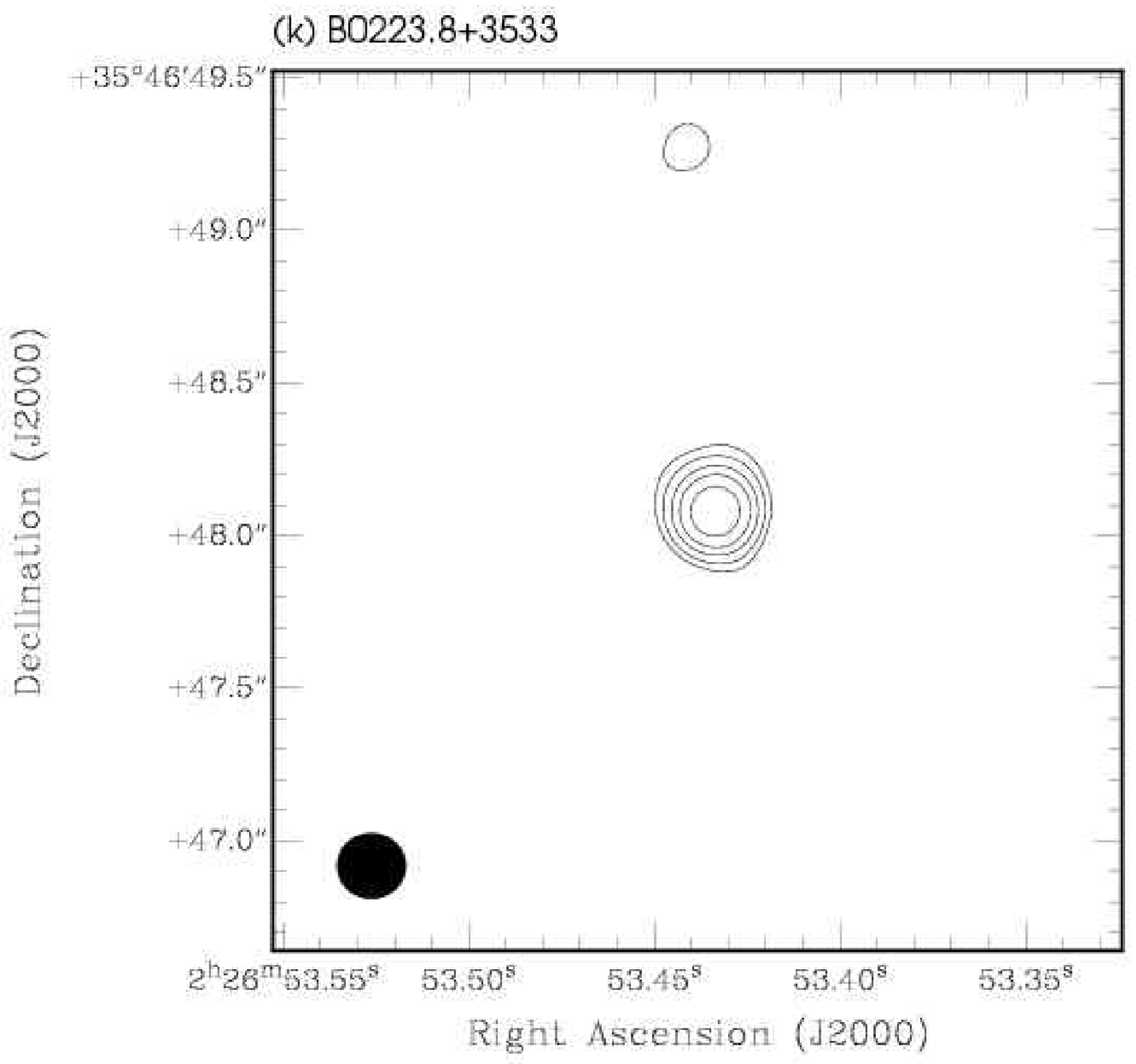} \quad
\plotone{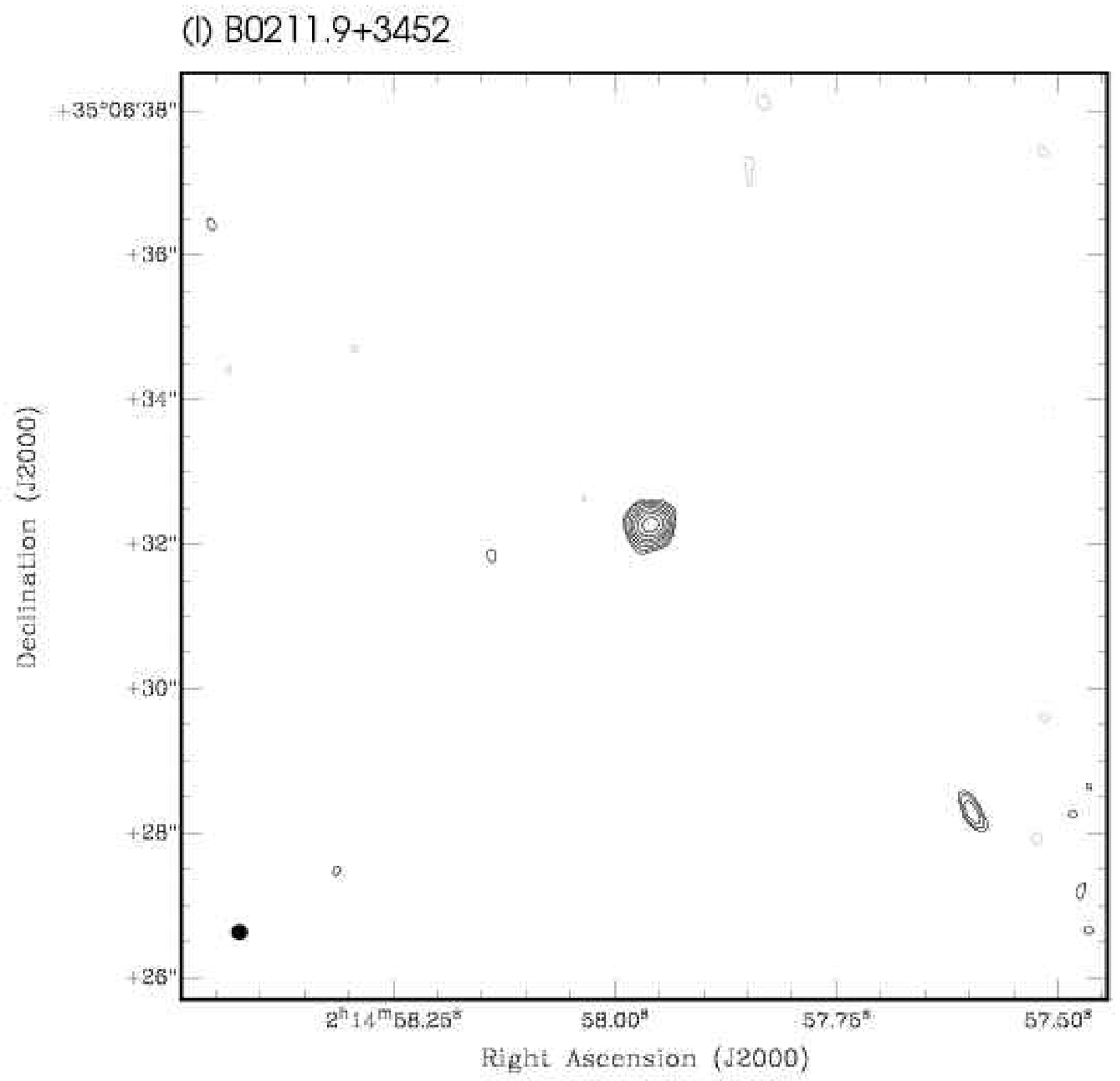}
}\\[5mm]
\mbox{
\plotone{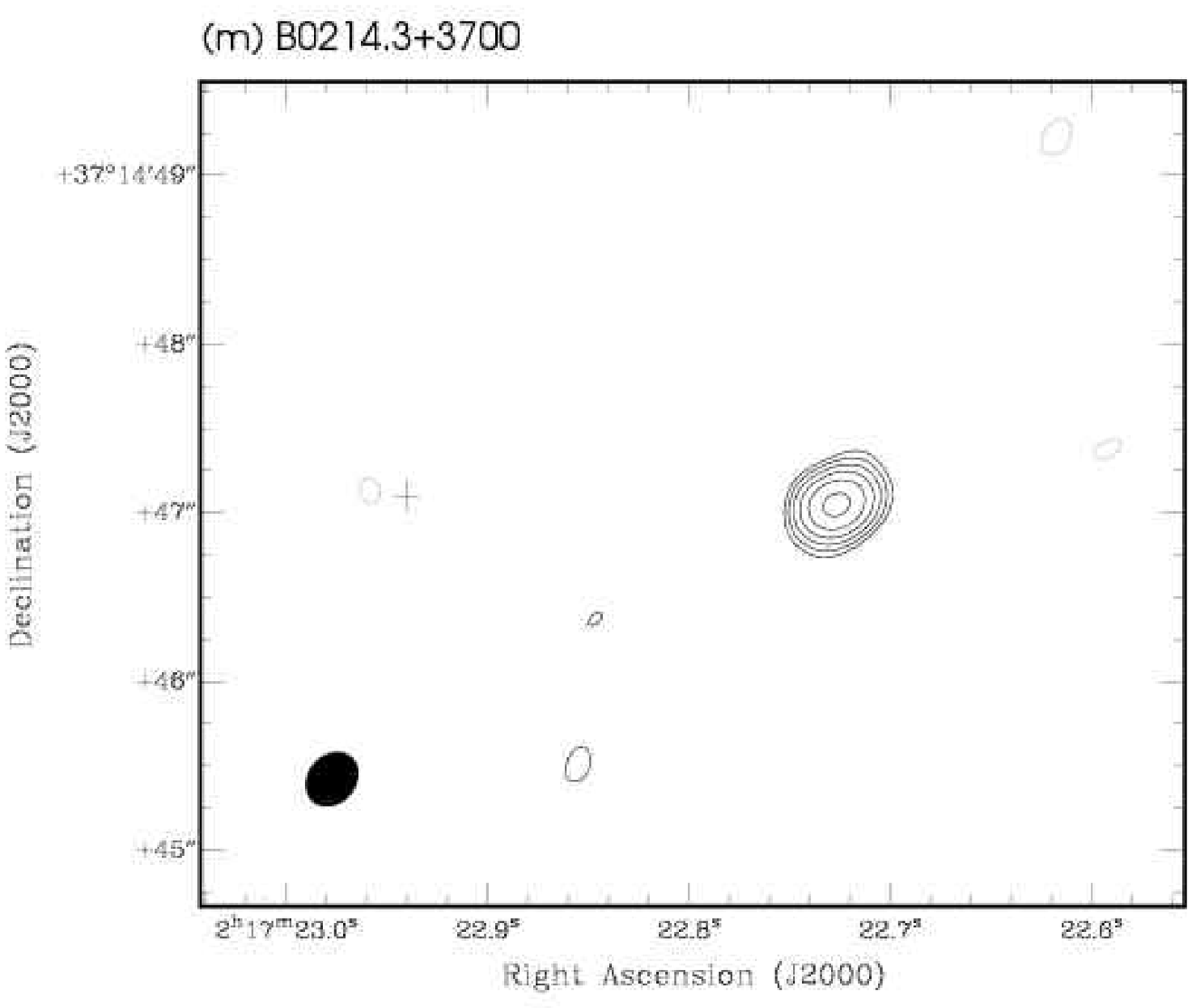} \quad
\plotone{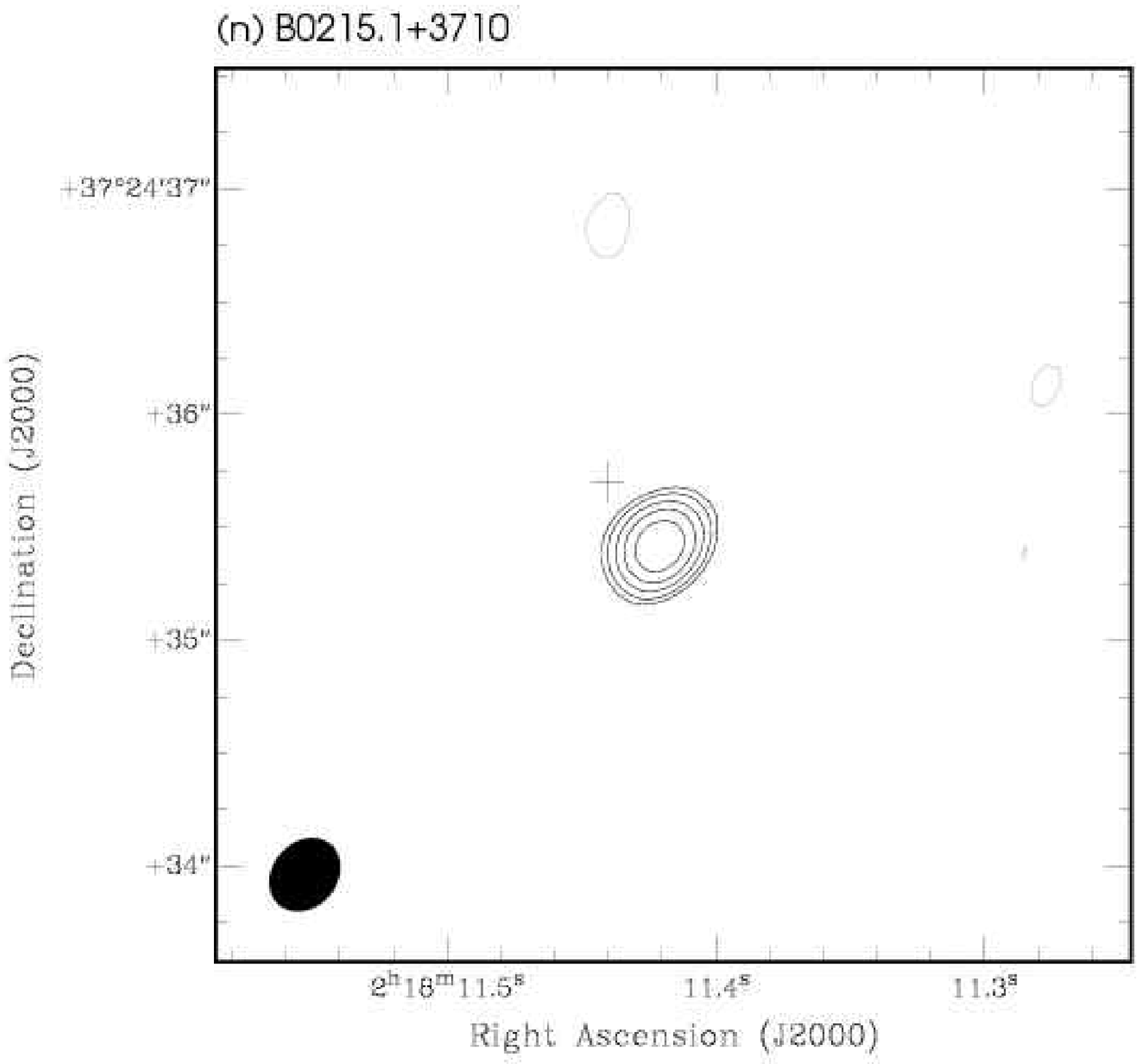}
}\\[5mm]
\mbox{
\plotone{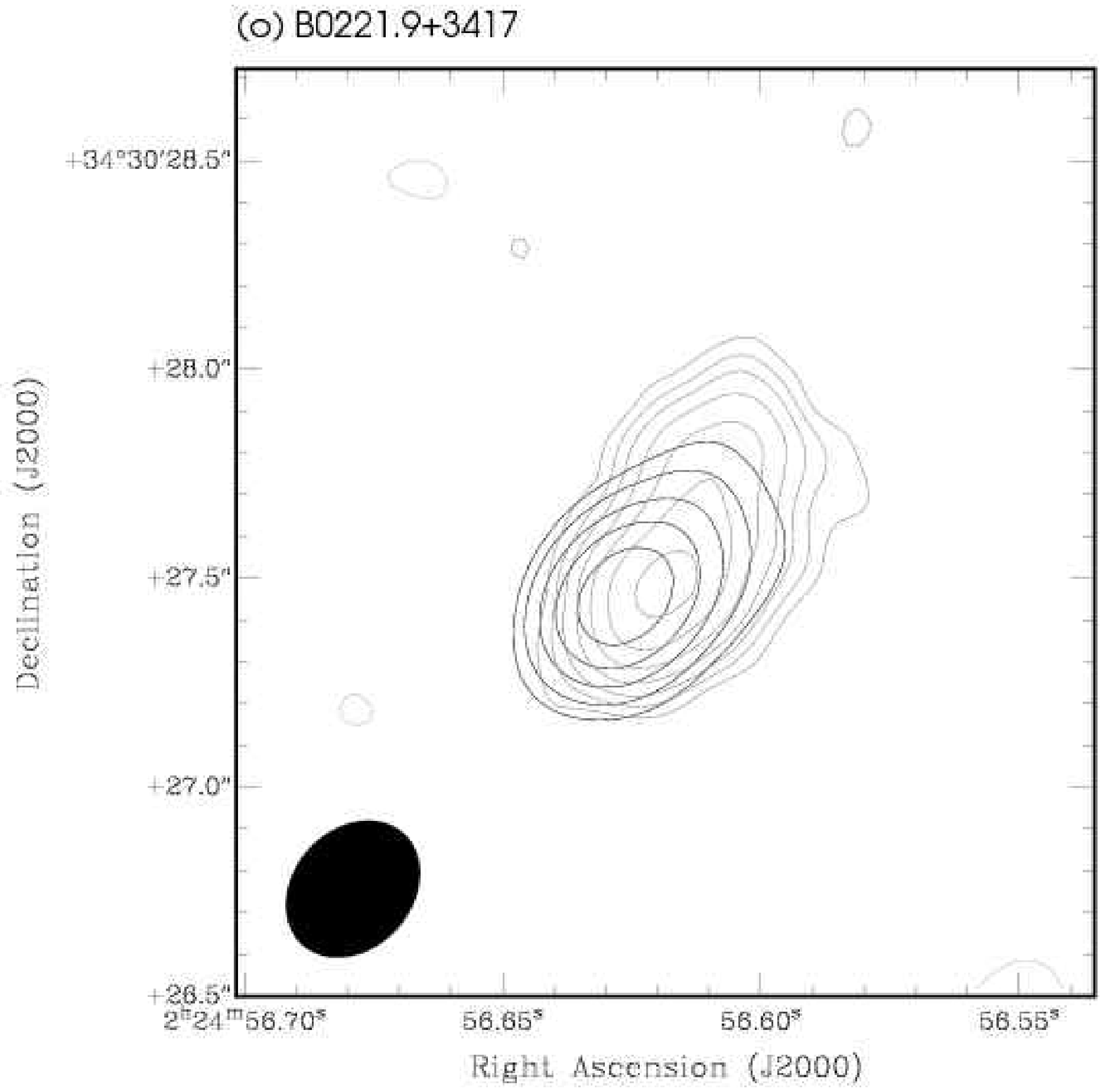} \quad
\plotone{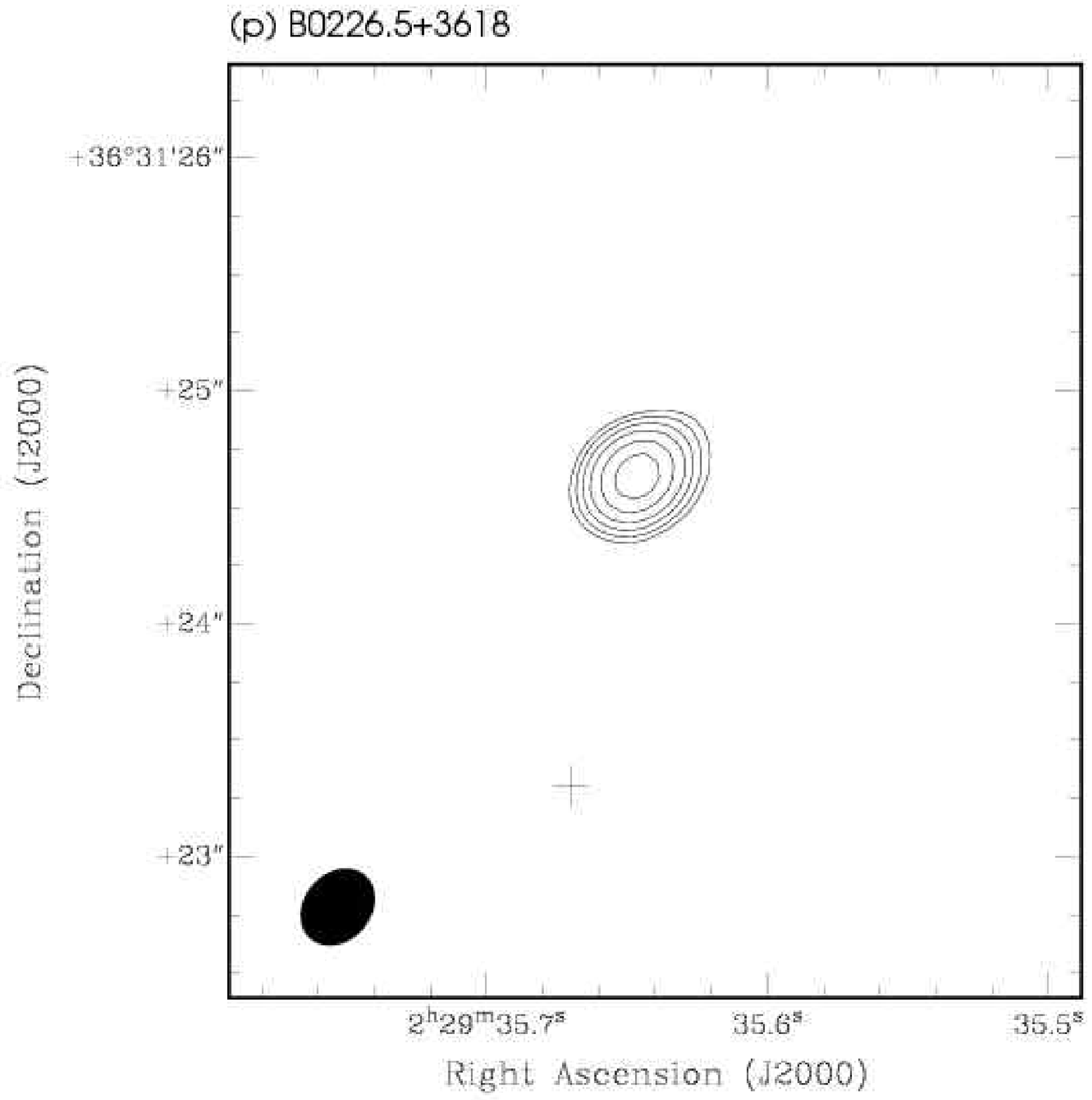}
}\\[5mm]
{Fig. 3. --- Continued}
\end{center}
\clearpage
\begin{center}
\mbox{
\plotone{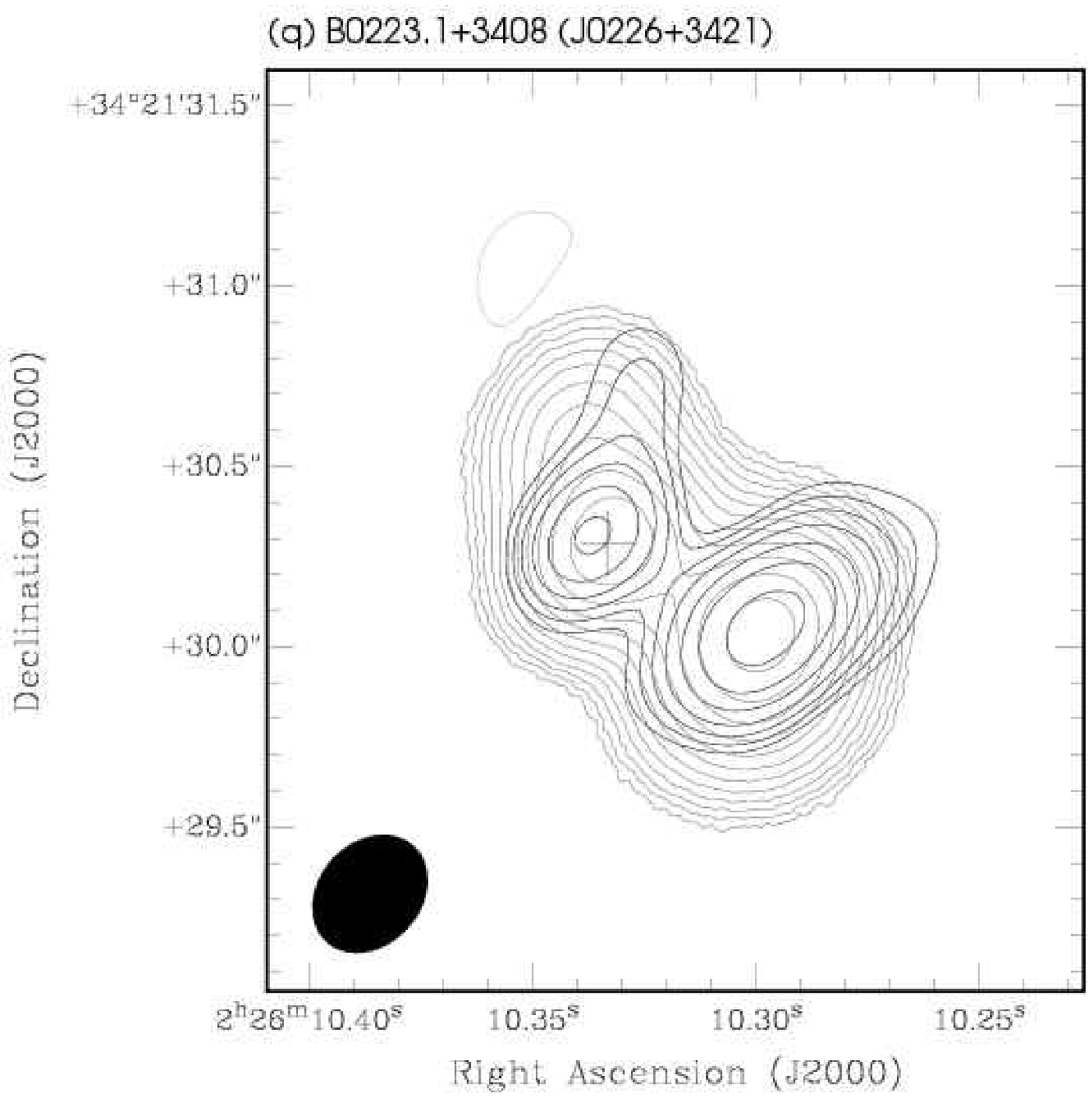} \quad
\plotone{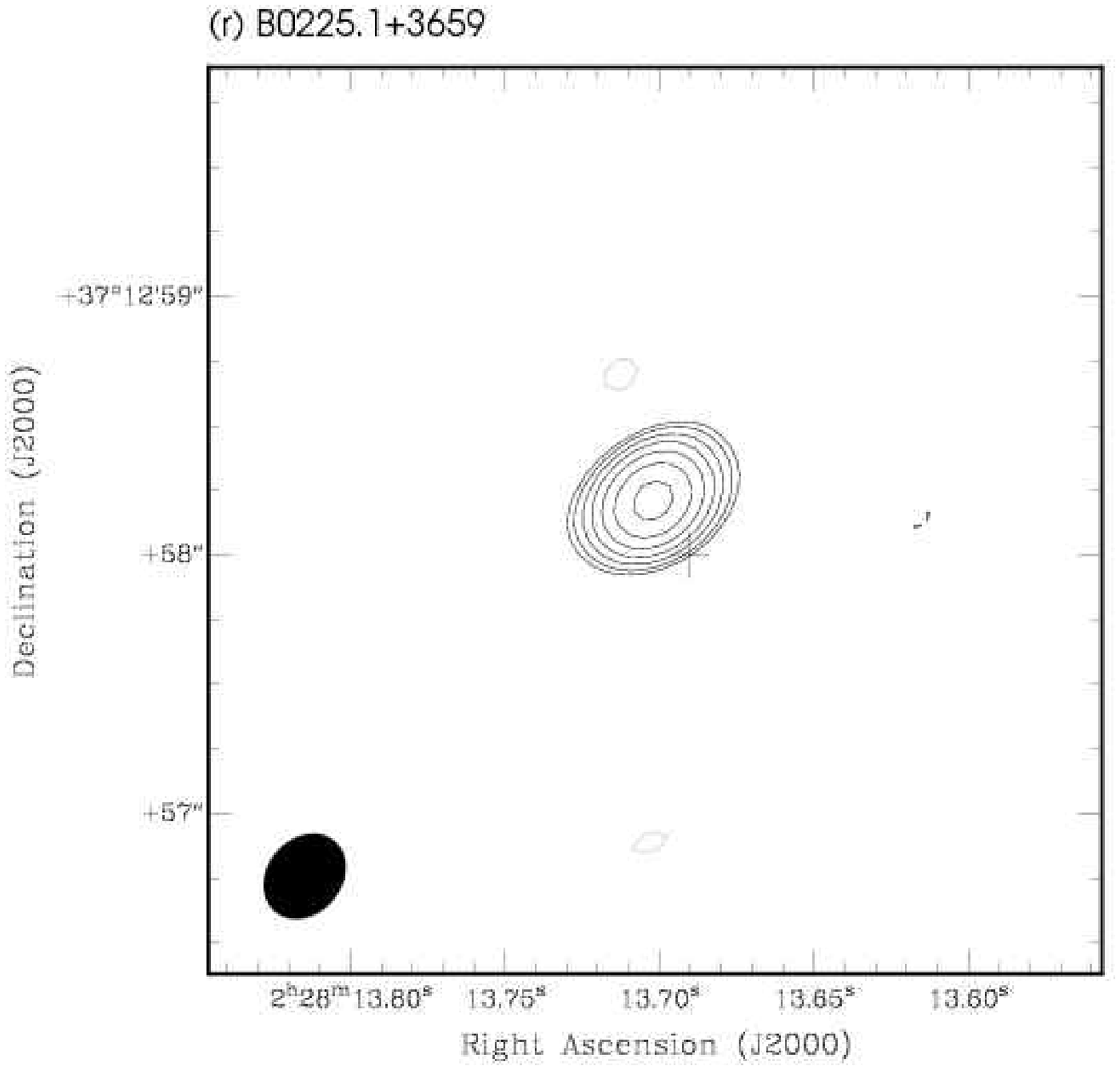}
}
\plotone{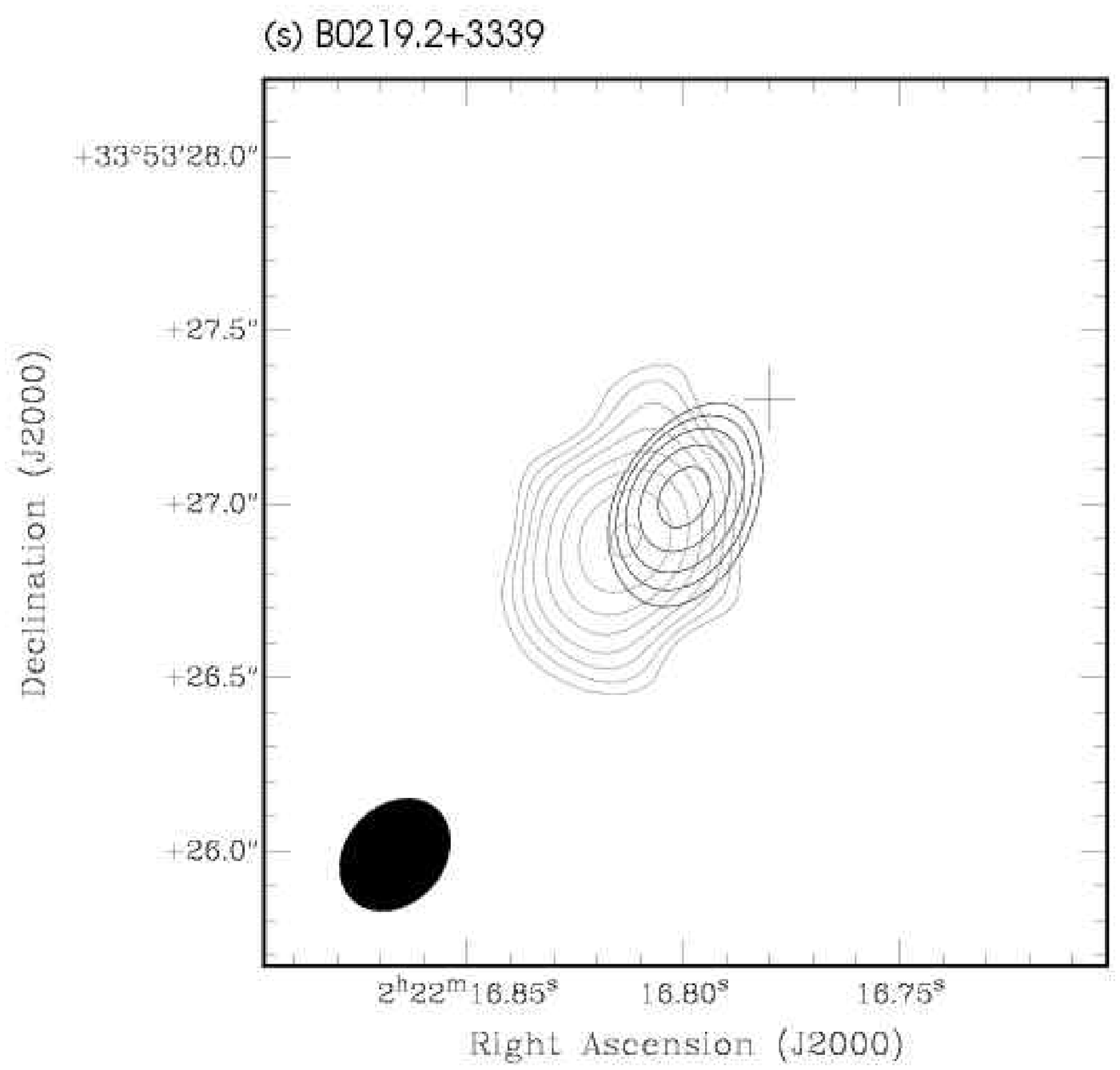}\\[5mm]
{Fig. 3. --- Continued}
\end{center}
\clearpage

\begin{figure}
\epsscale{0.4}
\plotone{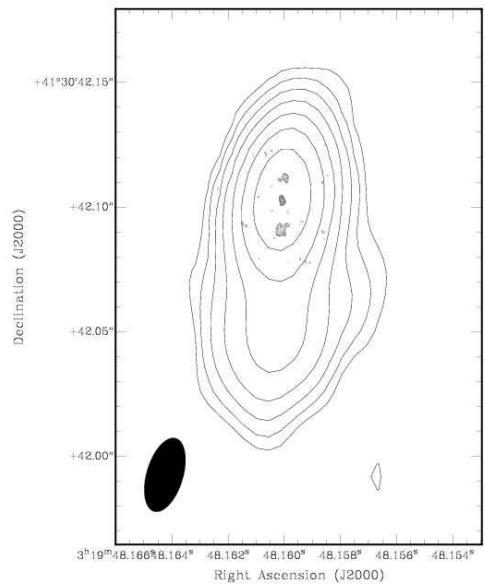}
\caption{Naturally weighted VLBI image of 3C84. Contours drawn at $\pm1, \pm2, \pm4, \cdots$ times the $3\sigma$ rms noise. Restoring beam and rms image noise for all images can be found in Table \ref{tab:tabastrometry}. A 15 GHz VLBA contour map is shown in grey \citep{lis05} with a peak flux density of 2.65 Jy beam$^{-1}$ and an integrated flux of 10.64 Jy. Contours are shown at 10, 20, 40, 80,$\cdots$, 2560 mJy beam$^{-1}$ and the restoring beam is $0.69\times0.55$ mas at a position angle of $2\arcdeg$.}
\label{fig:fig3c84}
\end{figure}

\clearpage

\end{document}